\documentclass[12pt]{article}
\usepackage{epsfig}
\usepackage{graphicx}
\usepackage{subfigure}
\usepackage{amssymb}
\usepackage{pifont}
\usepackage{latexsym}

\usepackage{amsmath}
\usepackage{amsfonts}
\usepackage{mathrsfs}
\usepackage{fancyhdr}
\usepackage{amscd}

\title{New insights into mode behaviours in waveguides with impedance boundary conditions}
\author{WenPing Bi, Vincent Pagneux}
\begin{document}

\maketitle
\begin{center}
Laboratoire d'Acoustique de l'Universit\'e du Maine,\\
UMR CNRS 6613\\
Av. O Messiaen, 72085 LE MANS Cedex 9,\\
France\\
\end{center}
%\affiliation{Laboratoire d'Acoustique de l'Universit\'e du Maine, UMR CNRS 6613, \\Facult\'e des sciences, 72085 Le Mans, France}%


 % Define commands to assure consistent treatment throughout document
% \newcommand{\eqnref}[1]{(\ref{#1})}
% \newcommand{\class}[1]{\texttt{#1}}
% \newcommand{\package}[1]{\texttt{#1}}
% \newcommand{\file}[1]{\texttt{#1}}
% \newcommand{\BibTeX}{\textsc{Bib}\TeX}
%

%
\begin{abstract}

In this paper we investigate mode nonorthogonal properties and their effects on the sound power attenuation in a waveguide with impedance boundary conditions. By introducing two quantities: self-nonorthogonality $K_p$, which measures the nonorthogonality between left and right eigenfunctions of a mode, and mutual-nonorthogonality $S_{ij}$, which measures the nonorthogonality between modes $i$ and $j$, two opposite limiting cases are clearly identified in the boundary impedance $\mathbb Z$ plane: one is non-dissipation, i.e., acoustic rigid, pressure-release, and purely reactive impedance; the other is Cremer's optimum impedances which are exceptional points --- a subject has attracted much attention in recent years in different physical domains. Variations along an arbitrary path in the complex boundary impedance plane, $K_p$ and $S_{i,j}$ varies between the two opposite extremes. It is found that $K_p$ and $S_{i,j}$ play crucial roles in sound power attenuation.   
\end{abstract}

%\pacs{43.20,Fn, 43.20,Hq, 43.20,Mv}% PACS, the Physics and Astronomy

%\maketitle

%\clearpage

% main text
\section{Introduction}

Modes in an infinite Waveguide with Impedance Boundary Conditions (WIBC) are a basic concept in acoustic textbooks, such as Refs. \cite{pierce} and\; \cite{morse}, and a powerful tool to understand the complex sound field in applications, such as ducts lined with locally-reacting acoustically absorbent materials, for review articles in aircraft engine duct systems see, for example, Refs. \cite{nayfeh} and the references therein. Mode method in a WIBC has been the subject of much research for more than 50 years. It is remarkable that there remain fundamental open questions; e.g., how to measure the nonorthogonality between modes when the boundary impedance is complex, and what are their effects on sound power attenuation?

%When the boundary conditions are acoustic rigid (Neumann boundary conditions), pressure-release (Dirichlet boundary conditions), and impedance but with only the reactive part, linear and lossless sound propagation in fluid is assumed, the problem is non-dissipation, Hermitian, and self-adjoint. The eigenvalues of modes are pure real or imaginary, the eigenfunctions of modes form a complete and orthogonal basis. The solutions of sound pressure (or particle velocity) can be expressed in terms of the eigenmodes. The total sound power can be uniquely expressed as the sum of the sound power of each mode. 

%When the boundary conditions are complex impedance, the problem is dissipative, non-Hermitian, and non-self-adjoint. The eigenvalues are complex, the eigenfunctions are bi-orthogonal or nonorthogonal. The nonorthogonality has been known from long time ago\cite{morse}. However, it has been understood only as a mathematical normalisation condition. Very little was known about its effects on sound propagation in a WIBC.

Another fundamental problem which is not full understood is the Cremer's optimum impedance. The optimization defined by Cremer\cite{cremer} is relative to the modal axial sound attenuation rates in an infinite WIBC. The maximum sound attenuation rate of individual mode is achieved by choosing the corresponding optimum wall impedance. Cremer\cite{cremer} investigated only the least attenuation mode. Tester\cite{tester1} generalised this concept to arbitrary higher order modes. Cremer's optimum impedance has been one of the most important liner design method\cite{tester1, eversman, mechel1, mechel2, zorumski, eversman1, koch, rice0, rice1, rice2, watson, watson1, campos, bielak}. Its optimum condition leads to double eigenvalues of the dispersion equation. These double eigenvalues of dispersion equation have been inquired by Morse\cite{morse1}, and studied by Tester\cite{tester1}, Zorumski\cite{zorumski}, Mechel \cite{mechel1, mechel2}, and Shendrov\cite{shenderov}. They form square-root branch points in the complex admittance plane\cite{tester1, mechel1, mechel2}. However, nothing is known about the eigenfunction behaviors in the vicinity of the branch points or optimum impedances up to now. 

Due to the nonorthogonality, total sound power is no longer the sum of sound power in individual modes, when multimodes propagate in a WIBC. Cross-powers make contribution to the total sound power. Creamer's optimum impedance aimes only at the maximum attenuation of individual mode. No attempt has been made to investigate the cross-power. Little was known about what are the effects of source and impedance boundary conditions on the cross-power propagations.

In this paper, we study mode nonorthogonal properties and their effects on the sound power attenuation in a  WIBC. The paper is organised as follows. In Sec. \ref{sec II},  we show that Cremer's optimum impedances are exceptional points, at which not only eigenvalues but also the associated eigenfunctions coalesce, the left eigenfunctions and right eigenfunctions of the coalescent modes are orthogonal. We introduce two physical quantities: self-nonorthogonality $K_p$, to measure the nonorthogonality between left and right eigenfunctions of individual mode; and mutual-nonorthogonality $S_{ij}$, to measure the nonorthogonality between modes $i$ and $j$. Two opposite limiting cases: $K_p=1$, $S_{ij}=0$ and $K_p=\infty$, $S_{i,j}=1$ are clearly identified in the whole complex boundary impedance $\mathbb Z$ plane, correspond to: non-dissipation, i.e., acoustic rigid, pressure-release, and purely reactive impedance, and the Cremer's optimum impedances, respectively. Variations along an arbitrary path in the complex boundary impedance plane, $K_p$ varies between $1$ and $\infty$, and $S_{i,j}$ varies between $0$ and $1$. The roles of $K_p$ and $S_{ij}$ in sound power attenuation in a semi-infinite WIBC are illustrated in Sec. \ref{propagation}.

The model of the present paper is chosen to be a cylindrical waveguide with circular cross-section. Such model is the most common in practical applications. The extensions to rectangular or annular waveguides are straightforward. Flow effects will be considered in the further work.

\section{Mode behaviors}\label{sec II}

We consider an infinite long cylindrical waveguide, of uniform and circular cross section, having locally reactive impedance wall boundary conditions. The impedance is assumed uniform along axial and circumferential directions, respectively.  Linear and lossless sound propagation in air is assumed. With time dependence $\exp(j\omega t)$ omitted,
the eigenvalues $\gamma$ and eigenfunctions $\tilde{\phi}$ of modes satisfies the Laplacian eigenvalue problem
\begin{equation}\label{laplacian}
\nabla^2_\bot\tilde{\phi}_{mn}=-\gamma_{mn}^2\tilde{\phi}_{mn},
\end{equation}
where
\begin{equation*}
\nabla^2_\bot =\frac{1}{r}\frac{\partial}{\partial r}(r\frac{\partial}{\partial r})+\frac{1}{r^2}\frac{\partial^2}{\partial \theta^2},
\end{equation*}
with the boundary condition
\begin{equation}\label{bound0}
\frac{\partial\tilde{\phi}_{mn}}{\partial r}=Y\tilde{\phi}_{mn}, \; \;  \mathrm{at} \; \; r=1,
\end{equation}
where $m$ and $n$ refer to, respectively, the circumferential and radial mode indices. $Y=-jK \beta_0$. $\beta_0=1/Z_0$, where $Z_0$ and $\beta_0$ are wall boundary impedance and admittance, respectively. They are complex number. $K=\omega R /c_0$ refers to the dimensionless frequency, $R$ is the radius of the waveguide.
By assuming the solution
\begin{equation}
\tilde{\phi}_{mn}(r, \theta)=\frac{J_m(\gamma_{mn}r)}{J_m({\gamma_{mn}})}\left\{\begin{array}{l}\cos(m\theta)\\
\sin(m\theta),\end{array}\right.
\end{equation}
we obtain the dispersion equation for the eigenvalues
\begin{equation}\label{dispersion1}
\gamma_{mn}\frac{J'_m(\gamma_{mn})}{J_m(\gamma_{mn})} = Y.
\end{equation}

If we define an operator $\mathscr{L}=\nabla^2_\bot+\gamma_{mn}^2$, the eigenvalue problem defined by Eqs. (\ref{laplacian}) and (\ref{bound0}) can be rewritten as
\begin{equation}\label{laplacian1}
\mathscr{L}\tilde{\phi}_{mn}=0,
\end{equation}
with the boundary condution
\begin{equation}\label{bound01}
\mathscr{G}\tilde{\phi}_{mn}=0,\; \text{at}\; r=1,
\end{equation}
where $\mathscr{G}=\partial/\partial r-Y$. We introduce a function $\tilde{\varphi}$ to define the adjoint eigenvalue problem (see appendix A)
\begin{align}
\mathscr{L}^+\tilde{\varphi}_{mn}=0, \;\; \mathscr{G}^+\tilde{\varphi}_{mn}=0\; \text{at}\; r=1,
\end{align}
where $\mathscr{L}^+=\nabla^2_\bot+(\gamma_{mn}^2)^*$, $\mathscr{G}^+=\partial/\partial r-Y^*$.

We will call $\tilde{\phi}_{mn}$ right eigenfunctions and $\tilde{\varphi}_{mn}$ left eigenfunctions in the following sections.

\subsection{Basic behaviors}\label{basic}

When the boundary is acoustically rigid ($\beta_0=0$), pressure release ($\beta_0=\infty$), or purely reactive without dissipation ($\beta_0=jc$, $c$ is real), the eigenvalue problems defined by Eqs. (\ref{laplacian}) and (\ref{bound0}) or (\ref{laplacian1}) and (\ref{bound01}) are self-adjoint (see Appendix A), i.e., $\mathscr{L}^+=\mathscr{L}$ and $\mathscr{G}^+=\mathscr{G}$. Therefore, $\tilde{\phi}_{mn}=\tilde{\varphi}_{mn}$. The eigenfunctions $\tilde{\phi}_{mn}$ form a complete set of function and are mutual-orthogonal in the sense
\begin{equation}\label{orthogonal}
\int_s\tilde{\phi}_{mn}\tilde{\phi}_{m'n'}^*ds=\Lambda_{mn}\delta_{mm'}\delta_{nn'},
\end{equation}
where ``$^*$" refers to complex conjugate, $\Lambda_{mn}$ are normalized constants, $\delta$ is Kronecker delta function, $s$ is the cross section of waveguides. 

On the other hand, when the wall impedance is complex, i.e., dissipation is included, the eigenvalue problems defined in Eqs. (\ref{laplacian}) and (\ref{bound0}) or (\ref{laplacian1}) and (\ref{bound01}) are not self-adjoint (see Appendix A). However, it can be proved (see Appendix A) that the eigenfunctions $\tilde{\phi}_{mn}$ and their adjoint $\tilde{\varphi}_{mn}$ are orthogonal
\begin{equation}
\int_s\tilde{\phi}_{mn}\tilde{\varphi}^*_{m'n'}ds=\Lambda'_{mn}\delta_{mm'}\delta_{nn'}.
\end{equation} 
This means that the eigenfunctions $\tilde{\phi}_{mn}$ ($\tilde{\varphi}_{mn}$) are not mutual-orthogonal, $\int_s\tilde{\phi}_{mn}\tilde{\phi}_{m'n'}^*ds\ne 0$, when $m\ne m', n\ne n'$, but bi-orthogonal,
\begin{equation}\label{bi-orthogonal}
\int_s\tilde{\phi}_{mn}\tilde{\phi}_{m'n'}ds=\Lambda'_{mn}\delta_{mm'}\delta_{nn'},
\end{equation}
where we have used $\tilde{\phi}_{mn}=\tilde{\varphi}^*_{mn}$ (see Appendix A). It is noted that there is no complex conjugate operation on the eigenfunctions $\tilde{\phi}_{m'n'}$. 

Wether the eigenfunctions are orthogonal or bi-orthogonal, in this paper, the eigenfunctions $\tilde{\phi}_{mn}$ and $\tilde{\varphi}_{mn}$ are normalized as
\begin{equation}\label{normalize}
\phi_{mn}(r, \theta)=\frac{1}{\sqrt{\Lambda_{mn}}}\tilde{\phi}_{mn}(r, \theta), \;\;\;\; \varphi_{mn}(r, \theta)=\frac{1}{\sqrt{\Lambda_{mn}}}\tilde{\varphi}_{mn}(r, \theta)
\end{equation}
where $\Lambda_{mn}$ are defined in Eq. (\ref{orthogonal}).

\begin{figure}[h!]
\begin{center}
\includegraphics[scale=0.5]{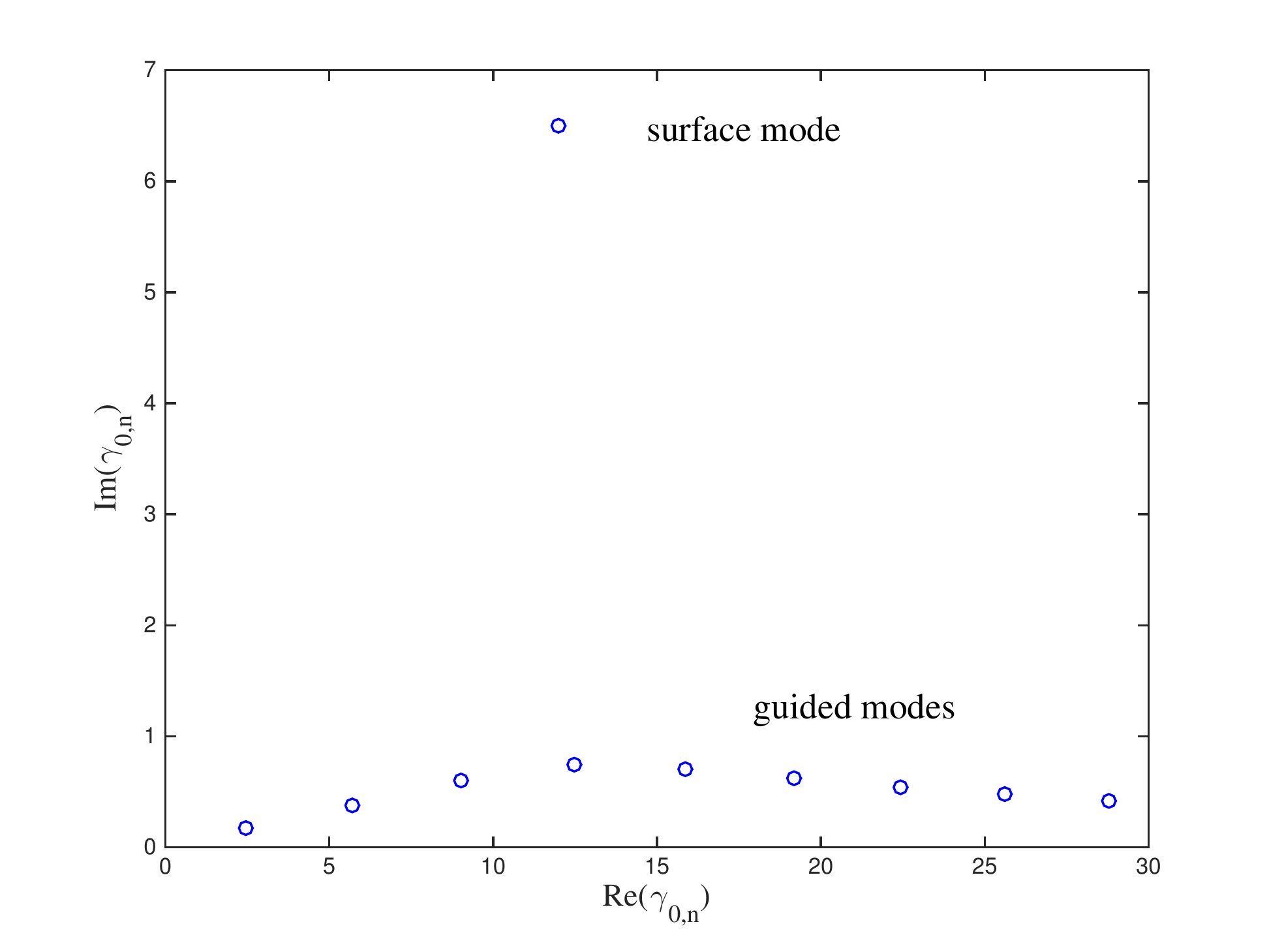}
\caption{\label{2D_m0} (Color online) Typical eigenvalues in a waveguide with impedance boundary conditions, $K=30$, $\beta_0=0.4+0.2j$, $m=0$.}
\end{center}
\end{figure}

\begin{figure}[h!]
\begin{center}
\includegraphics[scale=0.5]{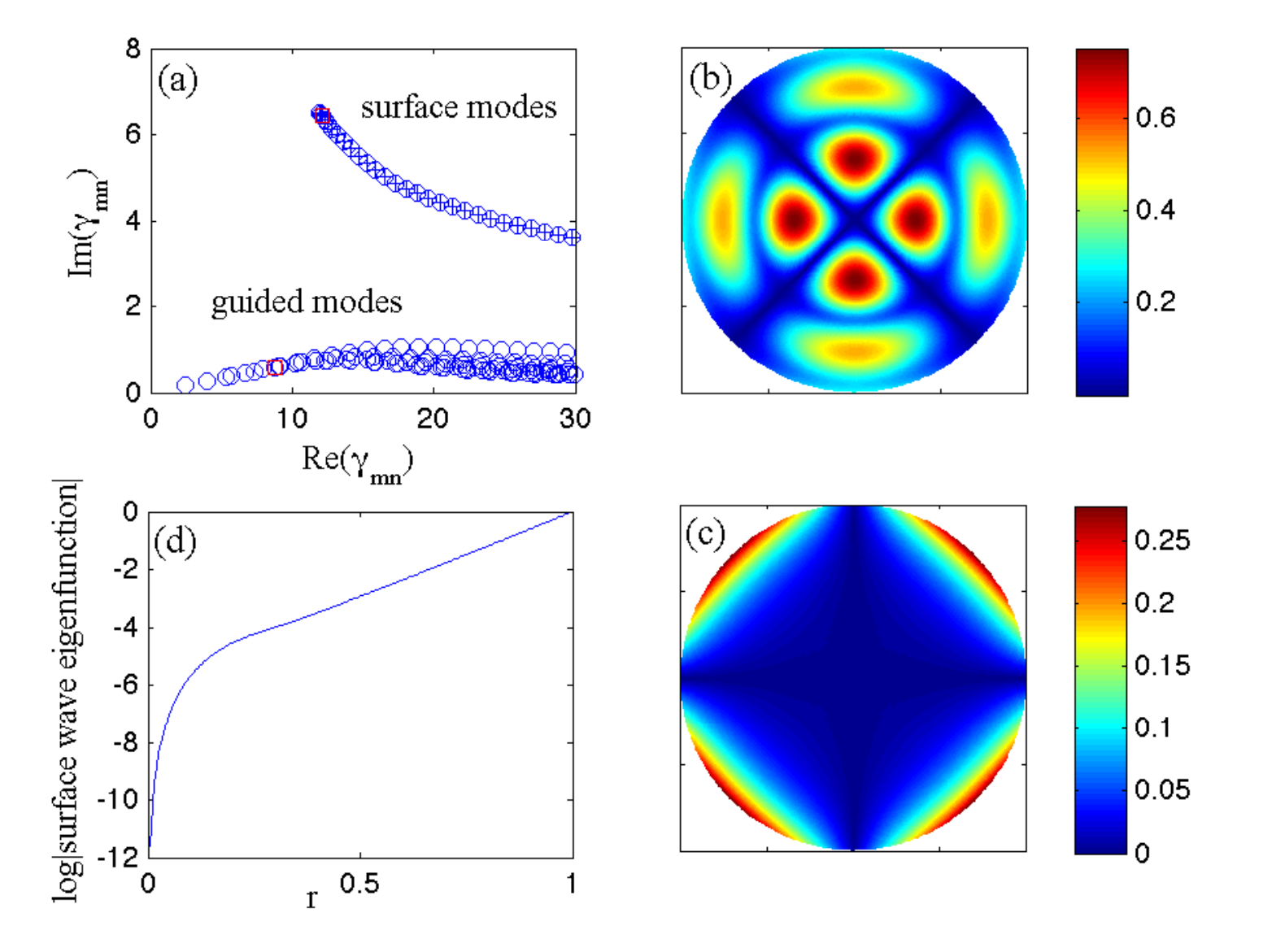}
\caption{\label{fig_md_uniform} (Color online) Eigenvalues and eigenfunctions of a WIBC, $K=30$, $\beta_0=0.4+0.2j$, $|m|=0-30$. (a) eigenvalues, $\oplus$ refers to surface modes (eigenvalues corresponding to Im$(\gamma_{mn})>3$ in this figure), (b) eigenfunction (not normalized) of guided mode (2,1), whose eigenvalue is shown as $\Box$ in the branch of guided modes in (a), (c) eigenfunction (not normalized) of surface mode $m=2$, whose eigenvalue is shown as $\Box$ in the branch of surface modes in (a), (d) the eigenfunction profile along $r$ of surface mode $m=2$.}
\end{center}
\end{figure}

There are an infinite number of modes in a WIBC corresponding to $m=0-\infty$ and $n=0-\infty$. They can be classified in two categories\cite{mechel1, mechel2, rienstra}: guided modes resulting from the finiteness of the waveguide geometry, and surface modes that exist only near the cavity wall and decay exponentially away from the wall. Typical eigenvalue distributions are shown in Fig. \ref{2D_m0} for $m=0$ and Fig. \ref{fig_md_uniform} for $|m|=0-30$, when $K=30$, $\beta_0=0.4+0.2j$ which are typical industrial values in the lined intakes of an aeroengine. There is only one surface mode when $m=0$ as shown in Fig. \ref{2D_m0}.
There are an infinite number of discrete surface modes in a WIBC corresponding to $m=0-\infty$, as shown in Fig. \ref{fig_md_uniform}(a) by ``$\oplus$". For each azimuthal order $|m|$ (except $m=0$), there are only two ($+|m|$ and $-|m|$) surface modes which are in degeneracy. It is noted that this degeneracy is totally different from the branch points and exceptional points in the following sections.  In Fig. \ref{fig_md_uniform} (a), each $\oplus$ corresponds to one $|m|$. They are arranged as $m=0, \;\; \pm 1, \;\; \pm 2, \;\;\cdots$, from left to right. The decaying rates of the surface mode amplitudes away from the wall are decided by the imaginary parts of the surface mode eigenvalues $\gamma_m$. A typical surface mode profile corresponding to $m=2$ is shown in Fig. \ref{fig_md_uniform}(c) and (d). It needs to stress that the surface modes in a WIBC are asymptotic solutions in high frequency $\omega$. The eigenfunctions become exponentially decaying along $r$ like $e^{\omega\vert\Im m(\gamma)\vert(1-r)}/\sqrt{r}$,\cite{rienstra} where $\Im$m refers to the imaginary part. Strictly speaking, they should be called  ``quasi-surface modes". The eigenvalues of guided modes are marked by ``o" in Fig. \ref{fig_md_uniform}(a). The eigenfunction of guided mode $(2,1)$, as an example, is plotted in Fig. \ref{fig_md_uniform}(b). 

Because the waveguide is circumferentially uniform. Modes among different azimuthal order $m$ are not coupled. In the following sections, we illustrate the results only for $m=0$. It is straightforward to extend the results to $m\ne 0$. The index $m=0$ is then omitted. Without loss of generality, we set $K=30$.

\subsection{Cremer's optimum impedance, branch points, and exceptional pointes}\label{EP}

Creamer's optimum impedance in an infinite WIBC has important applications in liner design to reduce noise in industry ducts. The optimization defined by Cremer is relative to the modal axial sound attenuation rates. The maximum sound attenuation rate of each mode is achieved by choosing the corresponding optimum wall impedance. Cremer's optimum condition leads to double eigenvalues of the dispersion equation as defined in Eqs. (\ref{double_root}). In the vicinity of the Creamer's optimum impedance, the eigenvalues, which have no power series expansion, are expressed approximately to the lowest order as\cite{tester1} (see Appendix B)
\begin{equation}\label{BP:expan}
\gamma_{n}-\gamma_{\text{cremer}}\approx-\sqrt{\frac{2\partial f/\partial\beta_0}{\partial^2 f/\partial\gamma^2_{n}}}\sqrt{\beta_0-\beta_{\text{cremer}}},
\end{equation} 
where we have assumed that the dispersion equation (\ref{dispersion1}) has no triple or higher order eigenvalues, $\beta_{\text{cremer}}$ refers to the admittance at Cremer's optimum impedance. Mathematically, Eq. (\ref{BP:expan}) clearly shows that Creamer's optimum impedance is a branch point in complex boundary impedance plane. 

\begin{figure}[h!]
\begin{center}
\includegraphics[scale=0.65]{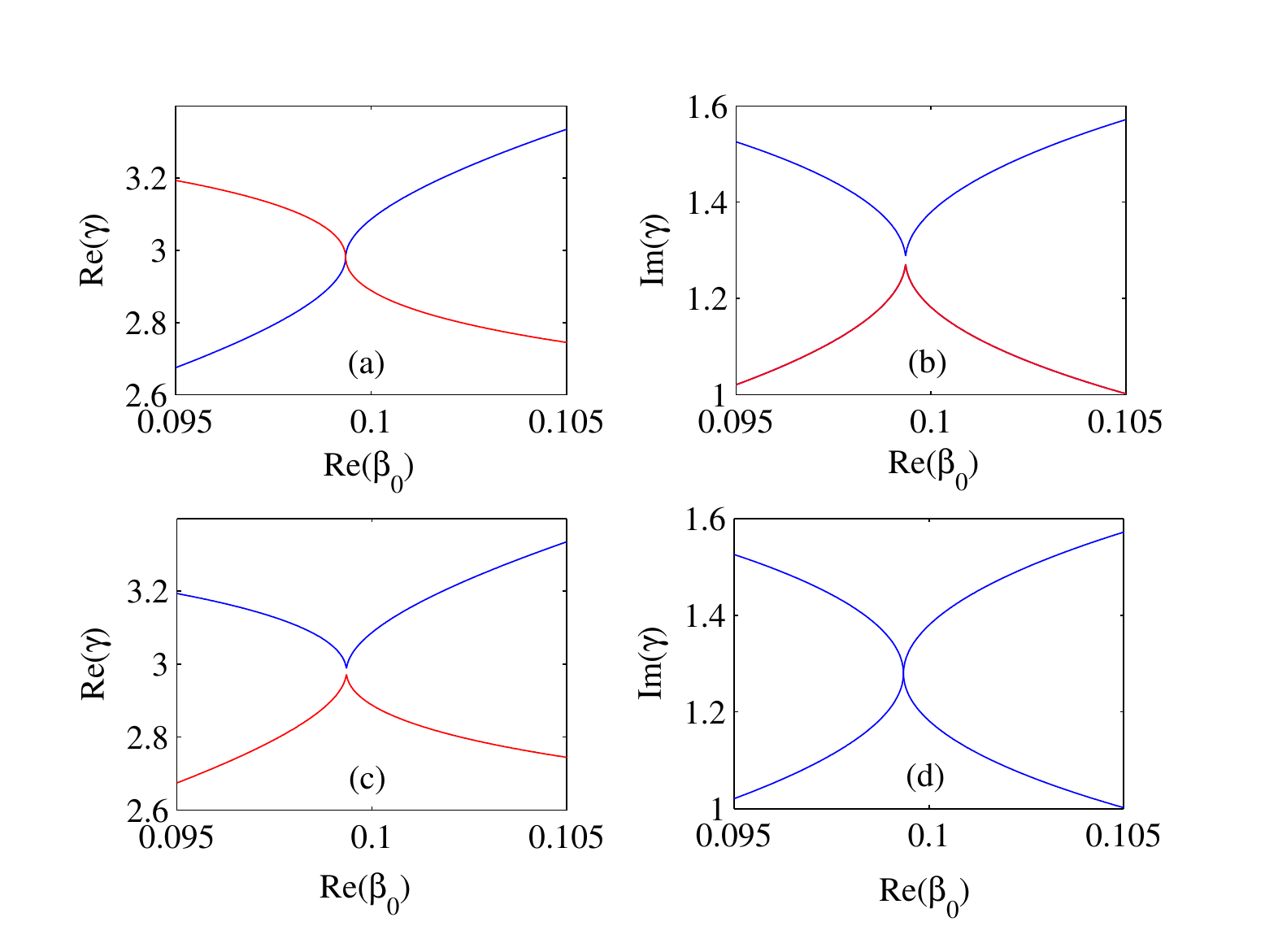}
\caption{\label{eig_beta_re}(Color online) Real and imaginary parts of eigenvalue trajectories of modes $n=0$ and $n=1$ in the vicinity of the first Cremer's optimum impedance as a function of $\Re e(\beta_0)$, $m=0$. $\beta_{\text{cremer}}=0.099346+0.042653j$. (a) and (b), $\Im m(\beta_0)=0.042655$; (c) and (d), $\Im m(\beta_0)=0.042652$.}
\end{center}
\end{figure}

The branch point behaviour can be proved to be a physical reality by an experiment. A numerical simulation is shown In Fig.\ref{eig_beta_re}. We plot the variations of the real and imaginary parts of eigenvalues as a function of $\Re e(\beta_0)$ when $\Im m(\beta_0)=0.042655$ ((a), (b)), and $\Im m(\beta_0)= 0.042652$ ((c), (d)) in the vicinity of the first Creamer's optimum impedance $\beta_{\text{cremer}}=0.099346+0.042653j$. The cusp (in Fig.  \ref{eig_beta_re} (b), (c)) originated from the square root behavior of the singularity is clearly seen. To illustrate the square root branch point singularity, we numerically encircle the Creamer's optimum impedance in the complex admittance plane in a complete loop: $(\Re e(\beta_0), \Im m(\beta_0))$$=(0.095, 0.042655)$$ - (0.105, 0.042655) - $$(0.105, 0.042652) - $\\ $(0.095, 0.042652) - $$(0.095, 0.042655)$. In this loop, the eigenvalues depend only weakly on $\Im m(\beta_0)$, we do not present the results for the varying $\Im m(\beta_0)$. After $\beta_{\text{cremer}}$ is encircled the complex eigenvalues are interchanged. It means that a full loop in the eigenvalue plane requires two loops in the complex admittance plane. A real experiment can re-produce the above process except that $\beta_0$ and $\gamma$ are less accurate. It is noted that the square root branch point behaviour has been experimentally observed by Dembowski \textit{et al}\cite{dembowski1} in a microwave cavity with dissipation, recently.
 
At the Creamer's optimum impedance, not only the eigenvalues of a pair of neighbour modes, but also the corresponding eigenfunctions coalesce. This can be illustrated by calculating the mutual-overlap integral for the mode pair $n$ and $n+1$ in the vicinity of Creamer's optimum impedance,
\begin{align}\label{eigfunc-overlap}
\int_s\phi_{n}(r, \theta)\phi^*_{n+1}(r, \theta)ds &=\int_s\frac{1}{\sqrt{\Lambda_{n}}}\tilde{\phi}_{n}(r, \theta)\frac{1}{\sqrt{\Lambda_{n+1}}}\tilde{\phi}^*_{n+1}(r, \theta)ds\\
& =\frac{\int_s\tilde{\phi}_{n}\tilde{\phi}^*_{n+1}ds}{\sqrt{\int_s\tilde{\phi}_{n}\tilde{\phi}^*_{n}ds\int_s\tilde{\phi}_{n+1}\tilde{\phi}_{n+1}^*ds}},\nonumber
\end{align} 
where we have used Eq. (\ref{normalize}).
\begin{figure}[h!]
\begin{center}
\includegraphics[scale=0.55]{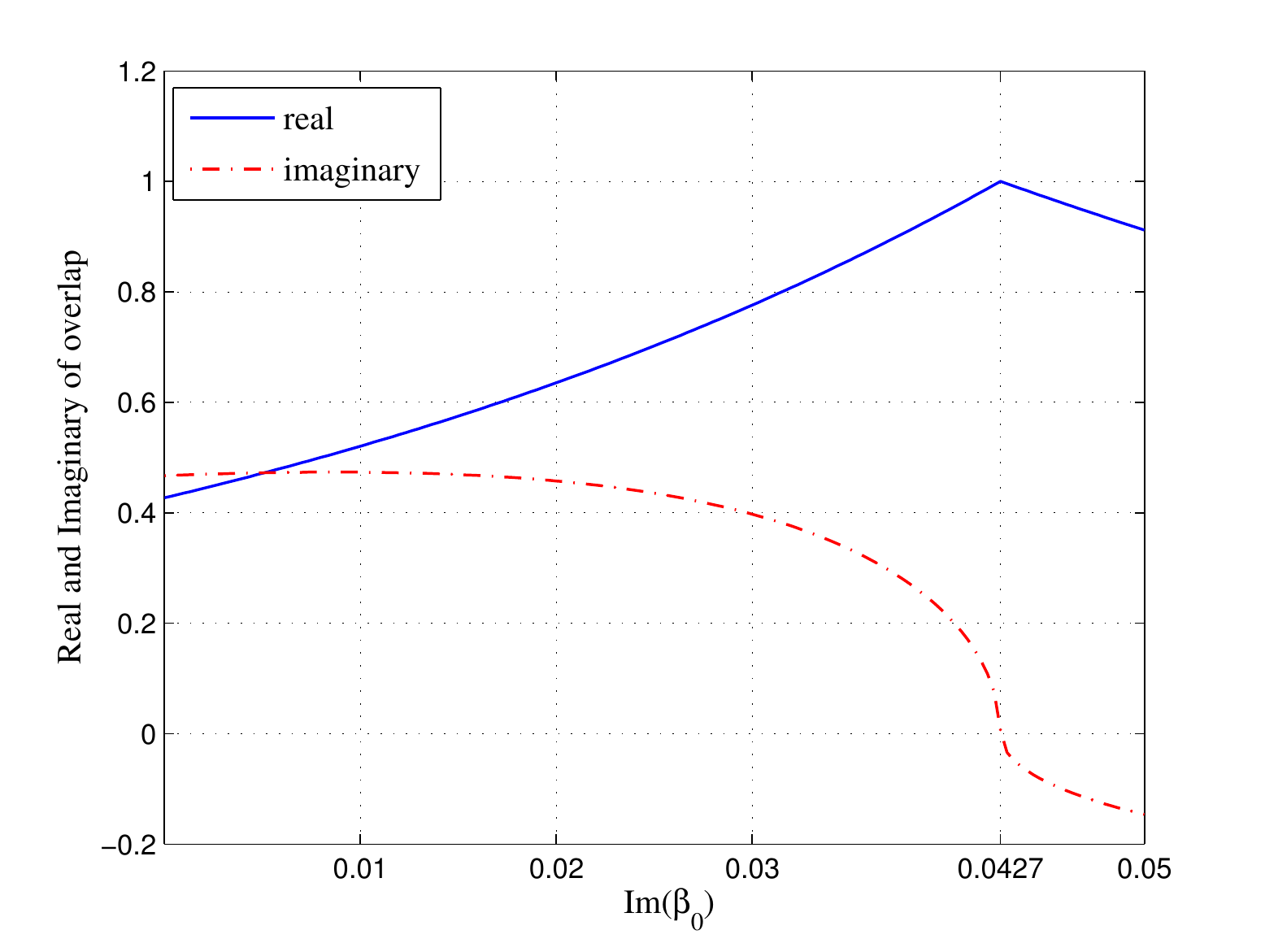}
\caption{\label{overlap_twomode} (Color online) Mutual-overlap integral Eq. (\ref{eigfunc-overlap}) of the eigenfunctions of mode pair $n=0$ and $n=1$ as a function of $\Im m(\beta_0)$, when $\Re e(\beta_0=0.099)$. $m=0$. Solid line, real part, dashed line, imaginary part. At Cremer's optimum impedance $\beta_{\text{creamer}} =0.099346+0.042653j$, the overlap integral is equal to 1.}
\end{center}
\end{figure}
In Fig. \ref{overlap_twomode}, we plot variations of the mutual-overlap integral of mode pair $n=0$ and $n=1$ as a function of $\Im m(\beta_0)$. The mutual-overlap integral is equal to 1 at $\beta_{\text{cremer}}=0.099346+0.042653j$. 

It can be further shown that at the Creamer's optimum impedance, the left and right eigenfunctions of the coalescent modes are orthogonal (self-orthogonality). This can be illustrated by calculating the self-overlap integral of the left and right eigenfunctions 
\begin{align}\label{lr-overlap}
\int_s\phi_{n}(r, \theta)\varphi^*_{n}(r, \theta)ds &=\int_s\frac{1}{\sqrt{\Lambda_{n}}}\tilde{\phi}_{n}(r, \theta)\frac{1}{\sqrt{\Lambda_{n}}}\tilde{\varphi}^*_{n}(r, \theta)ds\\
											    &=\frac{\int_s\tilde{\phi}_{n}\tilde{\varphi}^*_{n}ds}{\int_s\tilde{\phi}_{n}\tilde{\phi}^*_{n}ds}\nonumber.
\end{align}
\begin{figure}[h!]
\begin{center}
\includegraphics[scale=0.55]{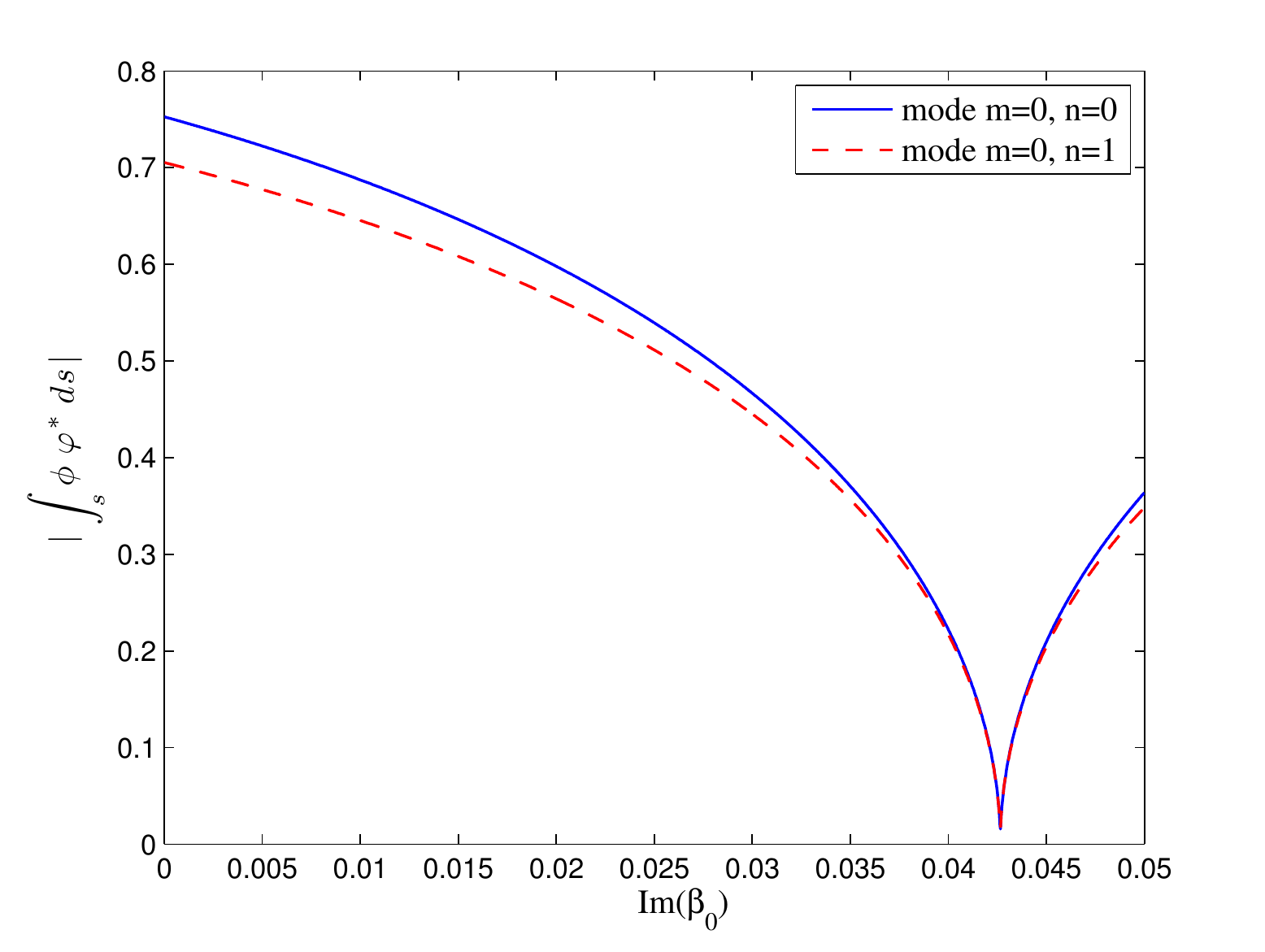}
\caption{\label{self_ortho_mode1} (Color online) Absolute value of the self-overlap integral Eq. (\ref{lr-overlap}) of the left and right eigenfunctions of modes $n=0$ and $n=1$ as a function of $\Im m(\beta_0)$, when $\Re e(\beta_0=0.099)$. Solid line, mode $n=0$, dashed line, mode $n=1$. $m=0$. At Cremer's optimum impedance $\beta_{\text{creamer}}=0.099346+0.042653j$, the self-overlap integral is equal to 0.}
\end{center}
\end{figure}
as a function of $\Im m(\beta_0)$ for mode $n$ in the vicinity of Creamer's optimum impedance. In Fig.\ref{self_ortho_mode1}, we plot variations of the absolute value of the self-overlap integral for modes $n=0$ and $n=1$. At the Cremer's optimum impedance $\beta_{\text{cremer}}=0.099346+0.042653j$, the self-overlap integral is equal to 0.

The point in a complex plane at which both eigenvalues and the corresponding eigenfunctions coalesce is called exceptional point (EP). EP should not be confused with a degeneracy, as mentioned above for the surface modes of $+\vert m\vert$ and $-\vert m\vert$, at which the corresponding eigenfunctions are still orthogonal. Recently, EPs have attracted much attention. The important properties of EPs have been uncovered by Heiss\cite{heiss1, heiss2, heiss3, heiss4}, Rotter\cite{rotter}, and Berry\cite{berry} for physical systems with dissipation or non-Hermitian system. EPs have been found in different systems, such as, laser-induced ionization states of atoms \cite{latinne}, electronic circuits \cite{Stehmann}, atoms in cross magnetic and electric fields \cite{cartarius}, a chaotic optical microcavity\cite{lee}, and $\mathscr{P}\mathscr{T}$-symmetric waveguides\cite{moiseyev}. This is the first time that EPs and their effects are illustrated in acoustics, to the best of the authors' knowledge.

%A simple example given by Heiss\cite{heiss4} is that when a physical system is formulated by a non-Hermitian matrix 
% The mathematically topological structures of Riemann sheets at an EP, which are a square-root branch point of the coalescing eigenvalues and a fourth-order branch point of the coalescing eigenfunctions depending on a complex or two real parameters have been proved physical reality\cite{Dembowski1}. EPs have attracted much attentions in non-Hermitian Hamiltonian quantum and optical systems (see, e.g., Refs. \cite{heiss3, rotter} and the references therein). They have been found in different domains, laser-induced ionization states of atoms\cite{latinne}, atom waves in crystals of light\cite{Oberthaler}, electronic circuits\cite{Stehmann}, atoms in cross magnetic and electric fields\cite{cartarius} and in microwave billiards\cite{Dembowski2}. Experiments in laboratories have been carried out with resonances in microwave cavities\cite{Dembowski1, Dembowski2, Dembowski3}. 

\begin{figure}[!htb]
\begin{center}
\includegraphics[scale=0.55]{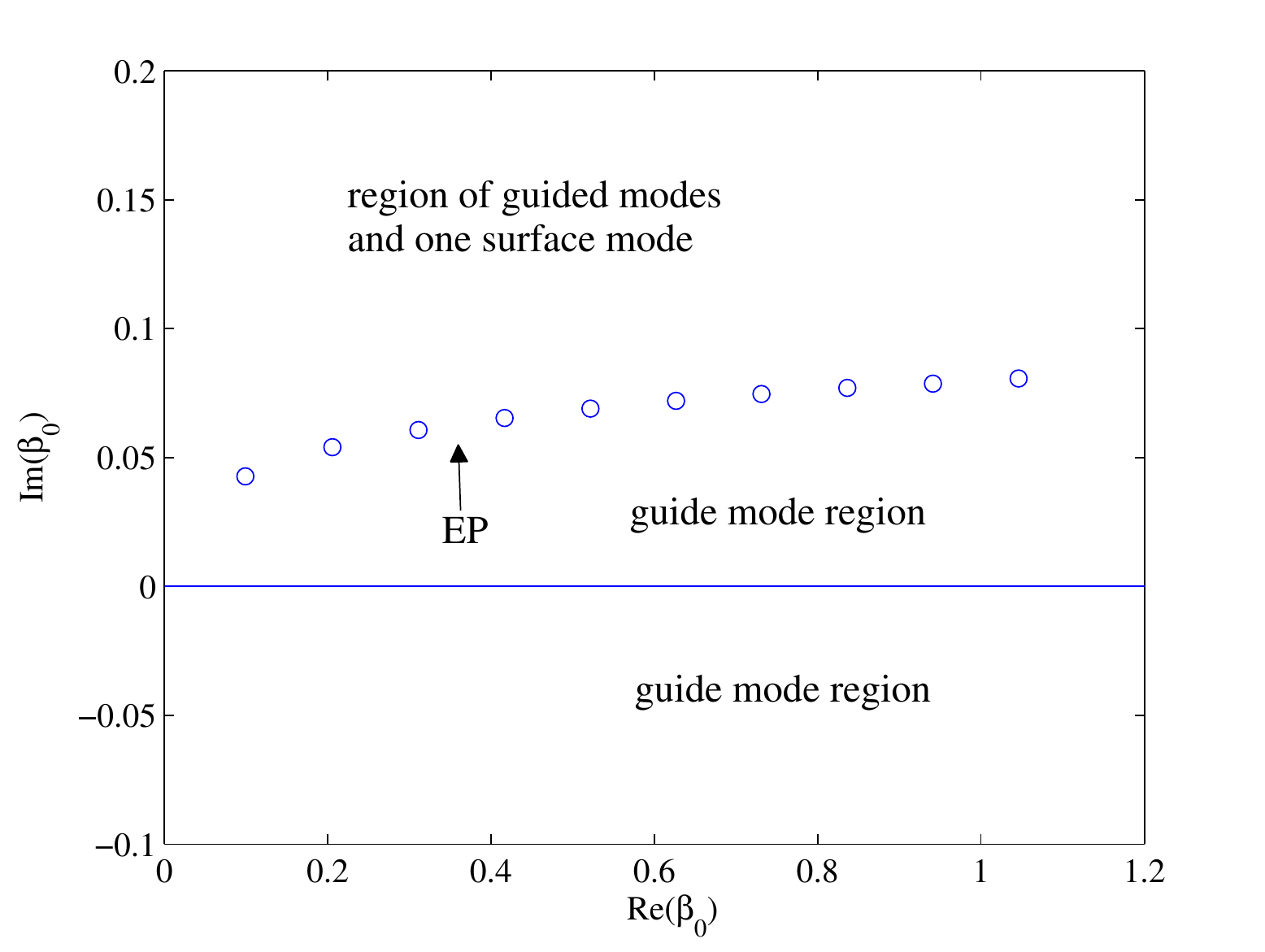}
\caption{\label{EP_distrib_m0} (Color online) Distribution of the first 10 EPs in the complex admittance plane, when $m=0$.}
\end{center}
\end{figure}
There are an infinite number of EPs in the complex admittance plane for each circumferenial index $m$. They can be calculated by Eq. (\ref{double_root}) (see Ref. \cite{mechel2}, for example). The first 10 EPs when $m=0$ are illustrated in Fig. \ref{EP_distrib_m0}. The EPs separate the complex admittance plane into two regions: in the lower region, there exist only guided modes, whereas in the upper region, there exist guided modes and one surface mode (for each $m$). The surface modes take place only in the $\Im m(\beta_0)>0$ plane (convention $e^{j\omega t}$ is used). 
%\begin{figure}[!htb]
%\includegraphics[scale=0.65]{eigen_vary_EP_im1.png}
%\caption{\label{eigen_vary_EP_im1} (Color online)  Eigenvalue trajectories passing near the first EP as a function of $\Im m(\beta_0)$. $\Re e(\beta_0)=0.09935>\Re e(\beta_{\text{EP}})$. $\Im m(\beta_0)=0 - 0.05$. `$\circ$', $\Im m(\beta_0)=0$; `$\square$', $\Im m(\beta_0)=0.05$; and `$\ast$' refers to near $\Im m(\beta_{\text{EP}})$. $m=0$. }
%\end{figure}

To finish this section, we would like to point out that the mechanism of Creamer's optimum impedance is not explained to date. As was pointed by Tester\cite{tester1} in 1973 that ``A most intriguing property of theoretical and experimental decay rates of modes in lined ducts, for which there is  no obvious explanation, is the existence of maximum decay rates for values of the liner impedance which, at first sight, are arbitrary and totally unconnected with any simple results associated with absorption by reflecting boundaries.''. This mechanism will be explained in Ref.\; \cite{bi_resonance_trap}.

\subsection{EPs, avoided crossings, and mode localisation}

Avoided crossings occur in the vicinity of an EP. This can be illustrated by a $2\times 2$ non-Hermitian matrix
\begin{equation}\label{matrixH1}
\mathsf{H}=\begin{pmatrix} \alpha_1 &  0\\ 0 & \alpha_2\end{pmatrix} + \lambda\begin{pmatrix} 0 &  c\\ c & 0\end{pmatrix},
\end{equation}
\begin{figure}[!htb]
\begin{center}
\includegraphics[scale=0.55]{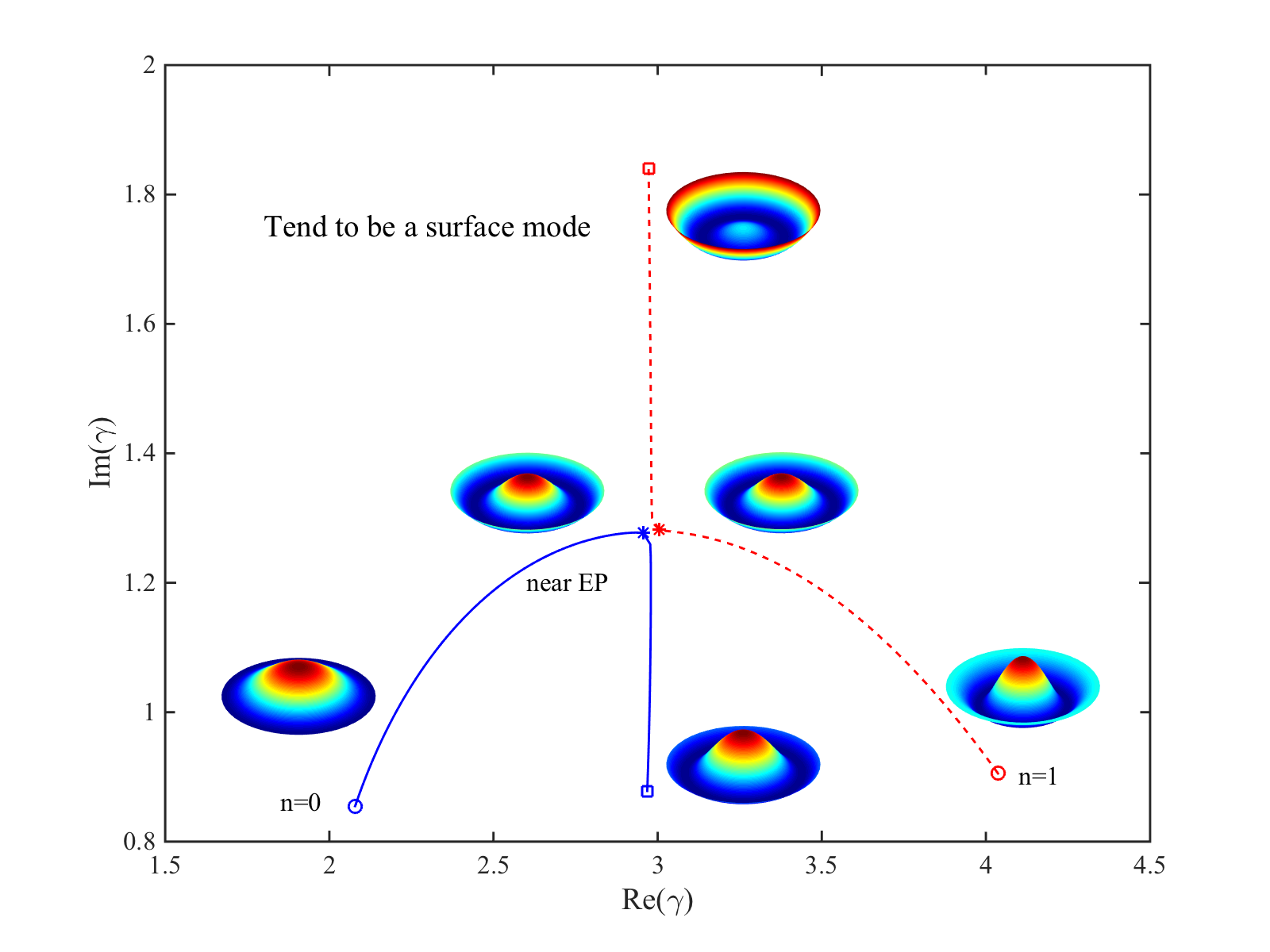}
\caption{\label{eig_vary_EP_im} (Color online)  Eigenvalue trajectories passing near the first EP as a function of $\Im m(\beta_0)$. $\Re e(\beta_0)=0.09935>\Re e(\beta_{\text{EP}})$. $\Im m(\beta_0)=0 - 0.05$. `$\circ$', $\Im m(\beta_0)=0$; `$\square$', $\Im m(\beta_0)=0.05$; and  `$\ast$' refers to near $\Im m(\beta_{\text{EP}})$. $m=0$. }
\end{center}
\end{figure}
%plot_eigf_uniform.m
where all the elements are complex. A corresponding physical problem can be found in Ref. \cite{bi_resonance_trap}. The eigenvalues of $\mathsf{H}$ are
\begin{equation}
\gamma_{1, 2}=\frac{1}{2}(\alpha_1+\alpha_2\pm R), \;\; R=\sqrt{(\alpha_1-\alpha_2)^2+4\lambda^2 c^2}).
\end{equation}
At the EPs, $\lambda_{\text{EPs}}=\pm(\alpha_1-\alpha_2)/(2c)$, $R=0$, the two eigenvalues coalesce $\gamma_{1, 2}=\frac{1}{2}(\alpha_1+\alpha_2)$, the corresponding eigenvectors also coalesce $\mathbf{x}_{1,2}=C[1, j]$ or $\mathbf{x}_{1,2}=C[1, -j]$. When $R\ne 0$, the eigenvalues avoided crossing as a function of $\lambda$. Avoided crossings of eigenvalues have been found in the area of structural dynamics\cite{leissa, kuttler, hodges, hodges1} and related to mode localisation in disordered structures\cite{pierre, pierre1, triantafyllou}. 

% \ref{eigen_vary_EP_im1}\ref{eigen_vary_EP_re} (in the sense that the phase of the  eigenfunction $-\pi/2<\text{arg}(\tilde{\phi}(r))<\pi/2$ or the real part of the eigenfunction has no node line)(in the sense that the phase of eigenfunction pass through $\text{arg}(\tilde{\phi}(r))=(2n+1)\pi/2$ one time, $n=0, 1, \cdots$, or the real part of the eigenfunction has one node line)
%In Fig.  \ref{eigen_vary_EP_re}, we plot the eigenvalue trajectories which pass near the second EP ($\beta_{\text{BP}}=0.205838+0.053957j$) as a function of $\Re e(\beta_0)$, when $\Im m(\beta_0)=0.054$ is fixed, and the corresponding eigenfunctions at some selected $\beta_0$, for $m=0$.  As $\Re e(\beta_0)$ {\em decrease}, the imaginary parts of eigenvalues of mode $n=0, \; 1, \; 2$ increase, and reach their maximum early or late. When $\Re e(\beta_0)$ approach the second EP, the eigenvalues of mode $n=1$ and $n=2$ form an avoided crossing and their eigenfunctions mix strongly. With a further {\em decrease} of $\Re e(\beta_0)$, mode $n=0$ and $n=1$ tend to mode $n=1$ and $n=2$ respectively, mode $n=3$ tend to a surface (localized) mode. 
%\begin{figure}[!htb]
%\includegraphics[scale=0.65]{eigen_vary_EP_re.png}
%\caption{\label{eigen_vary_EP_re} (Color online)  Eigenvalue trajectories passing near the second EP as a function of $\Re e(\beta_0)$. $\Im m(\beta_0)=0.054$, $\Re e(\beta_0)=0 - 1.667$. `$\circ$', $\Re e(\beta_0)=0.013$; `$\square$', $\Re e(\beta_0)=1.667$; and `$\ast$' refers to near $\Re e(\beta_{\text{EP}})$. $m=0$.}
%\end{figure}

The eigenvalue trajectories in the vicinity of the first EP ($\beta_{\text{BP}}=0.099346+0.042653j$) is shown in Fig. \ref{eig_vary_EP_im} as a function of $\Im m(\beta_0)$, when $\Re e(\beta_0)=0.09935>\Re e(\beta_{\text{EP}})$ is fixed. The eigenfunctions at some selected $\beta_0$ are also plotted. As $\Im m(\beta_0)$ increase, the imaginary parts of the eigenvalues of mode $n=0$ and those of mode $n=1$  increase until $\beta_0$ approaches the EP, where the eigenvalues form an avoided crossing and the eigenfunctions mix strongly. With a further increase of $\Im m(\beta_0)$, mode $n=1$ turns to be a surface mode which is localized near the guide wall as mentioned in section \ref{basic}, and mode $n=0$ turn to a mode which resembles mode $n=1$. The modes exhibit a similar behavior as we plot the eigenvalue trajectories as a function of $\Im m(\beta_0)$, when $\Re e(\beta_0)=0.09933<\Re e(\beta_{\text{EP}})$ is fixed. The only difference is that it is mode $n=0$ turn to be a surface mode and mode $n=1$ return to a mode which resembles mode $n=1$. 

It needs to stress that the mode localisation mentioned above is explained as a "resonance trapping" effect\cite{bi_resonance_trap}, and is different from these studied in Ref. \cite{pierre, pierre1, triantafyllou} which are due to disorder effects.

\subsection{Riemann surfaces}

\begin{figure}[htp]
\begin{center}
\includegraphics[scale=0.45]{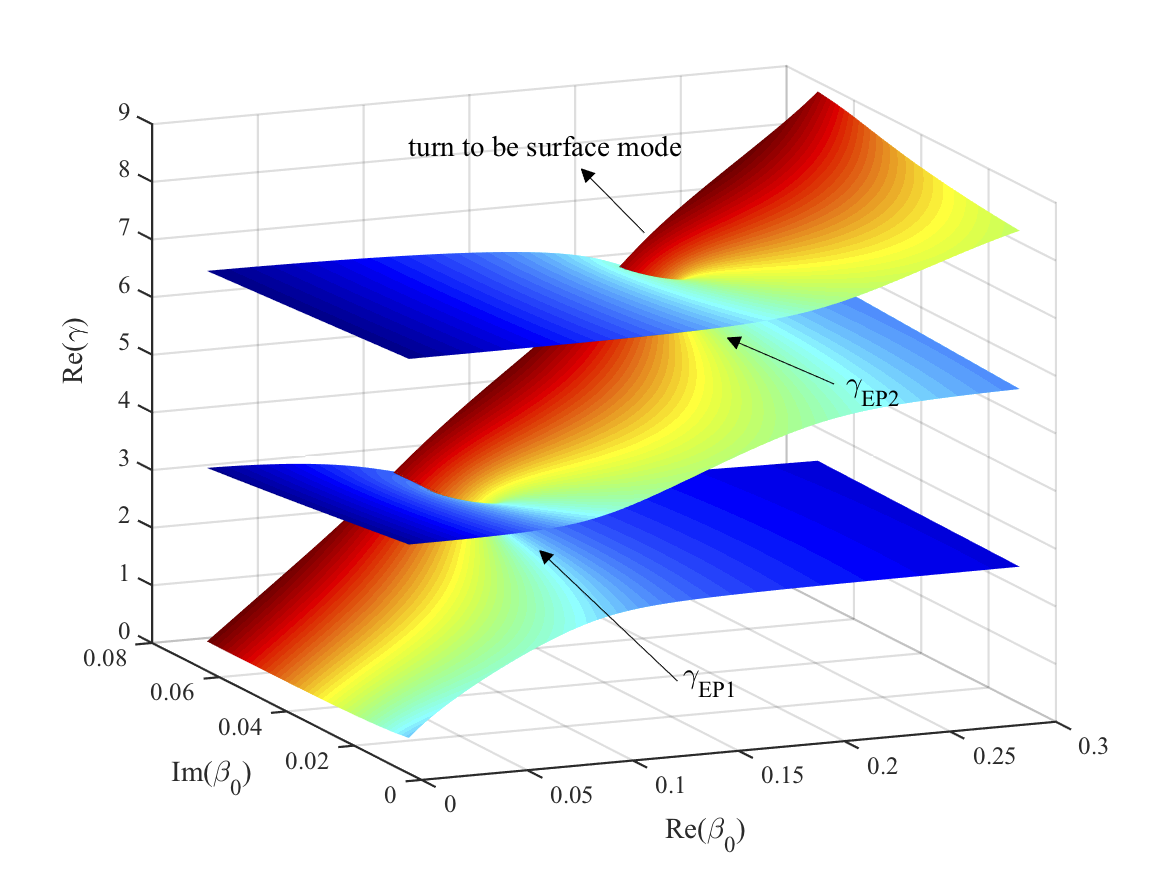}
\caption{\label{riemann_3D_re} (Color online) Riemann surfaces of the eigenvalues as a function of admittance $\beta_0$. Real part of the eigenvalues. The first two EPs are shown. $m=0$.}
\end{center}
\end{figure}

\begin{figure}[htp]
\begin{center}
\includegraphics[scale=0.45]{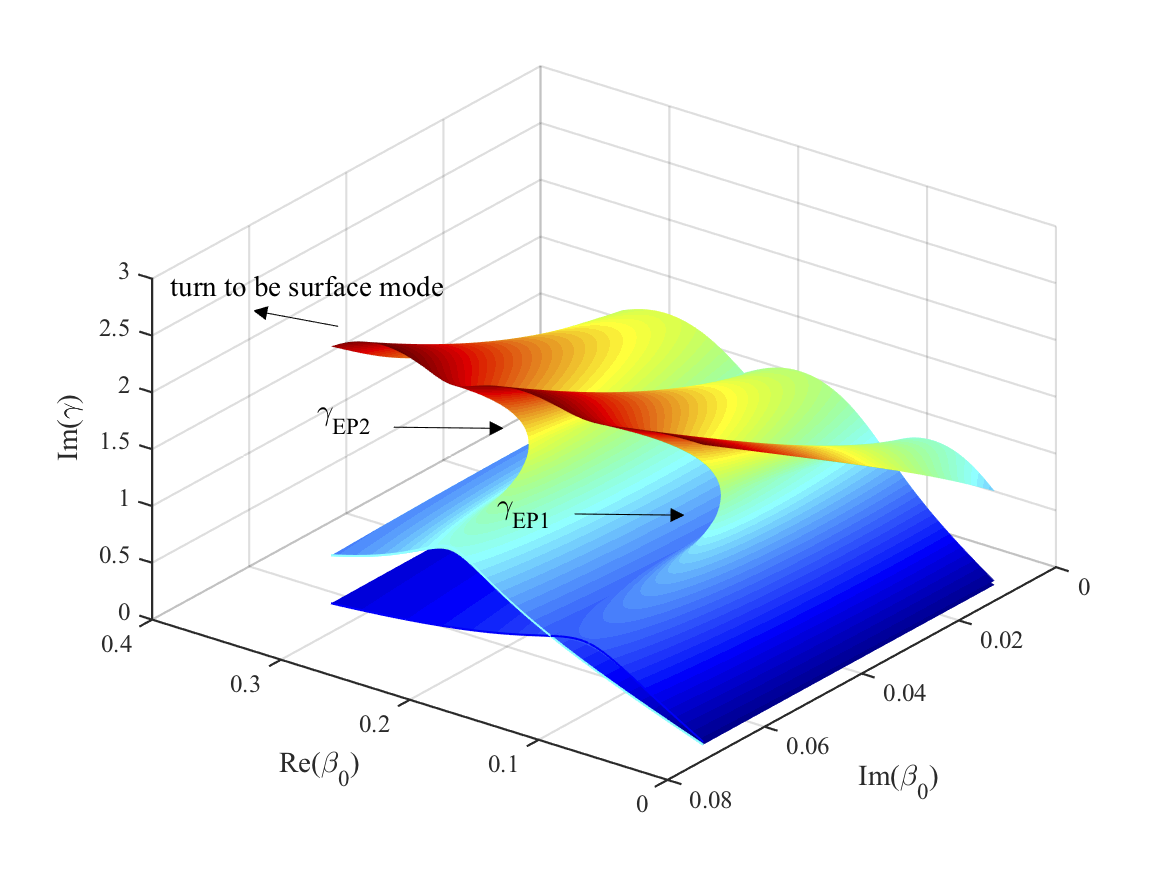}
\caption{\label{riemann_3D_im} (Color online) Riemann surfaces of the eigenvalues as functions of admittance $\beta_0$. Imaginary part of the eigenvalues. The first two EPs are shown. It is noted that to see more clearly the structures of the Riemann surfaces, the imaginary parts are shown in different viewpoint. $m=0$.}
\end{center}
\end{figure}

Another way to illustrate the structures of eigenvalues in the vicinities of EPs and the connections between EPs and avoided crossings is to plot the Riemann surface of eigenvalues over $\beta_0$ surface as shown in Figs. \ref{riemann_3D_re} and \ref{riemann_3D_im} for the first three modes. The values at the first two EPs $\gamma_{\text{EP1}}$ and $\gamma_{\text{EP2}}$ are pointed out in the figures. Modes with higher circumferential modal order ($m > 1$) or higher radial mode order ($n\ge 2$) exhibit similar characteristics.

It is well known that mode eigenvalues can vary continuously from one mode to another with continuous varying impedance. This can also be seen from the Riemann surfaces in Figs. \ref{riemann_3D_re} and \ref{riemann_3D_im}. The branch cuts separating one mode from another is arbitrary. When surface modes are present, it is not easy to find unambiguous branch cuts to distinguish the modes. In this paper, we define the mode index $n=0,\; 1, \cdots$, according to the ascending order of the real parts of eigenvalues.

\subsection{Self-nonorthogonality $K_{p}$ and mutual-nonorthogonality $S_{ij}$}\label{Petermann}

Inspired by the coalescence of eigenfunctions between two neighbour modes and orthogonality between left and right eigenfunctions of the coalescent modes at an EP, we define two quantities: self-nonorthogonality $K_{p,n}$ and mutual-nonorthogonality $S_{ij}$
\begin{equation}\label{kp}
K'_{p,n}=\frac{\sqrt{\int_s\tilde{\phi}_{n}\tilde{\phi}^*_{n}ds\int_s\tilde{\varphi}_{n}\tilde{\varphi}^*_{n}ds}}{\int_s\tilde{\phi}_{n}\tilde{\varphi}^*_{n}ds},\;\; K_{p,n}=K'_{p,n}*(K_{p,n}')^*,
\end{equation}
\begin{equation}\label{sij}
S_{ij} =\frac{\int_s\tilde{\phi}_{i}\tilde{\phi}^*_{j}ds}{\sqrt{\int_s\tilde{\phi}_{i}\tilde{\phi}^*_{i}ds\int_s\tilde{\phi}_{j}\tilde{\phi}_{j}^*ds}}=\int_s\phi_{i}(r, \theta)\phi^*_{j}(r, \theta)ds,
\end{equation}
to measure the nonorthogonality between left and right eigenfunctions of individual mode $n$ and nonorthogonality between modes $i$ and $j$ in the whole complex impedance (admittance) plane, respectively. It is noted that $K_p$ has been proposed by Petermann\cite{petermann} for explaining the discrepancy between the theoretically expected natural line-width using the Schawlow-Townes formula and the experimental measured enhanced line-width of a gain-guided single mode semiconductor laser.

Two opposite limiting cases can be identified in the complex $\beta_0$ plane:

 Case 1: The boundary conditions are non-dissipation, i.e., acoustic rigid, pressure-release, and purely reactive impedance. Modes are mutual-orthogonal, i.e., $\int_s\tilde{\phi}_{i}\tilde{\phi}^*_{j}ds=0$, left eigenfunctions $\tilde{\varphi}_n$ are equal to right eigenfunctions $\tilde{\phi}_n$. In this case, $S_{ij}=0$ and $K_p=1$. 
 
\begin{figure}[htp]
\begin{center}
\includegraphics[scale=0.45]{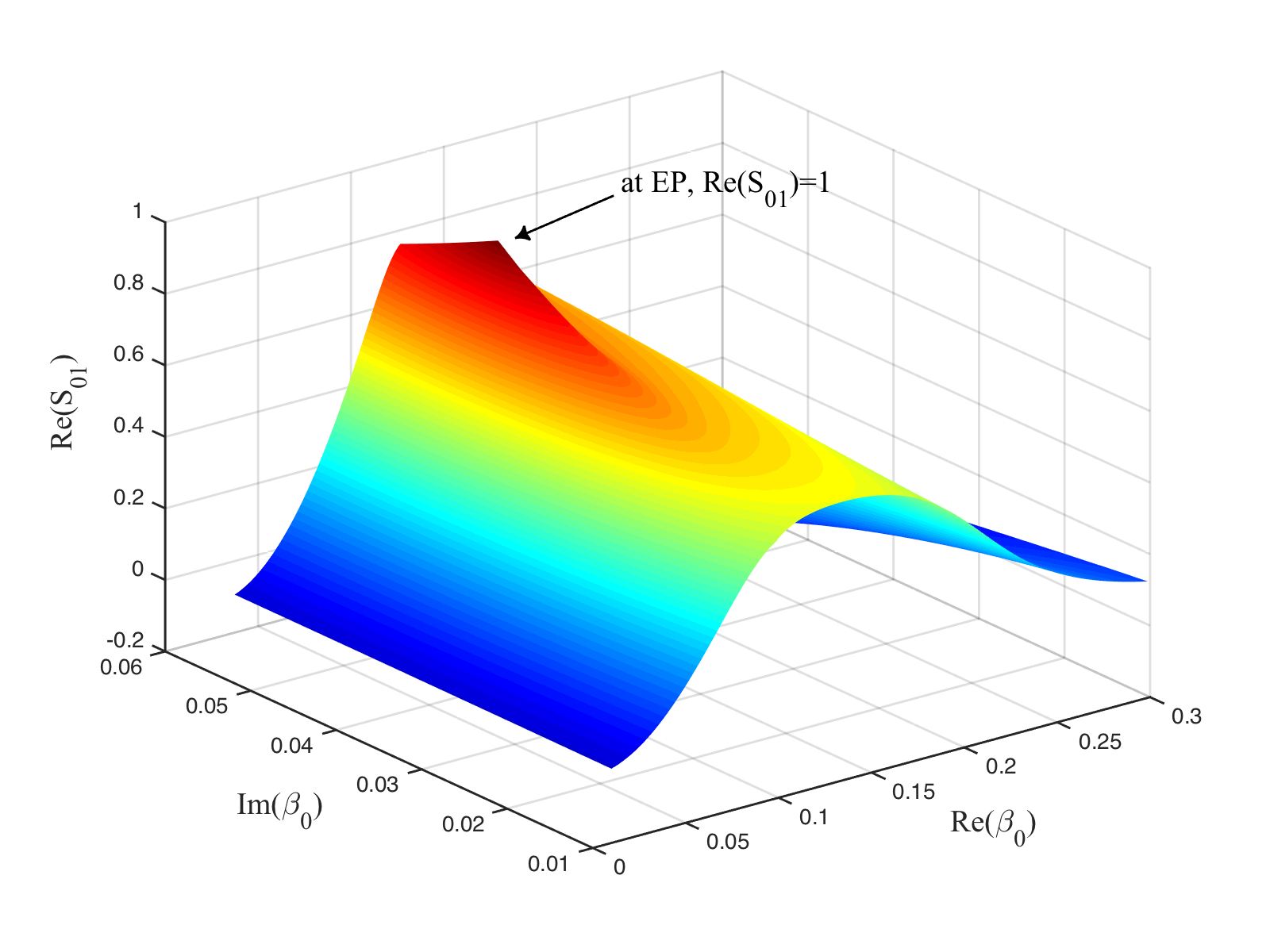}\hfill
\includegraphics[scale=0.45]{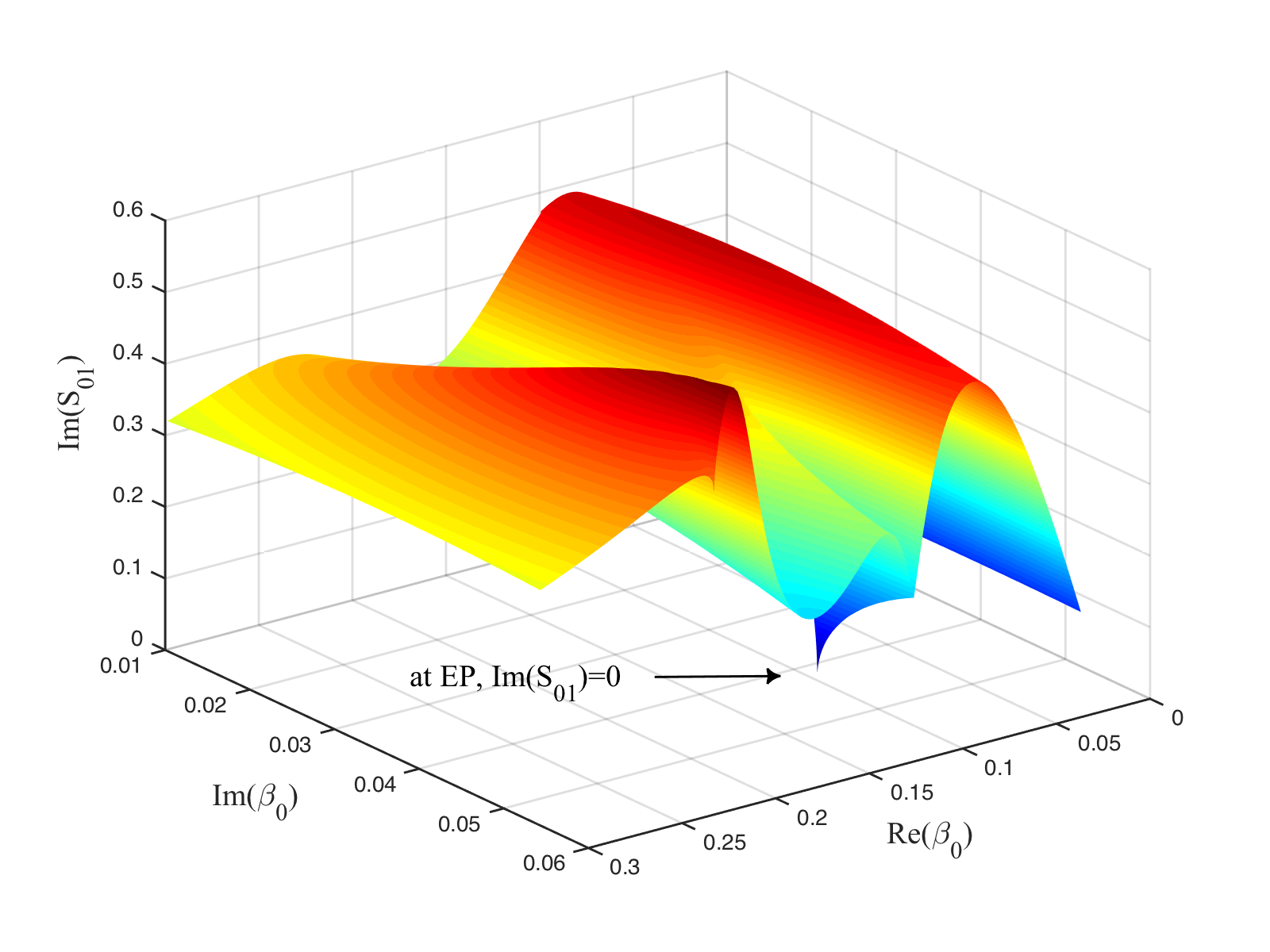}
\caption{\label{Sij_fig} (Color online) Real and imaginary parts of mutual-nonorthogonality $S_{01}$ over the complex $\beta_0$ plane. It is noted that for seeing clearly $\Im m(S_{01})=0$ at the first EP, we set 
x-axis ($\Re e(\beta_0)$) values decrease from left to right, and y-axis ($\Im m(\beta_0)$) values decrease from front to back.}
\end{center}
\end{figure}

\begin{figure}[htp]
\includegraphics[scale=0.6]{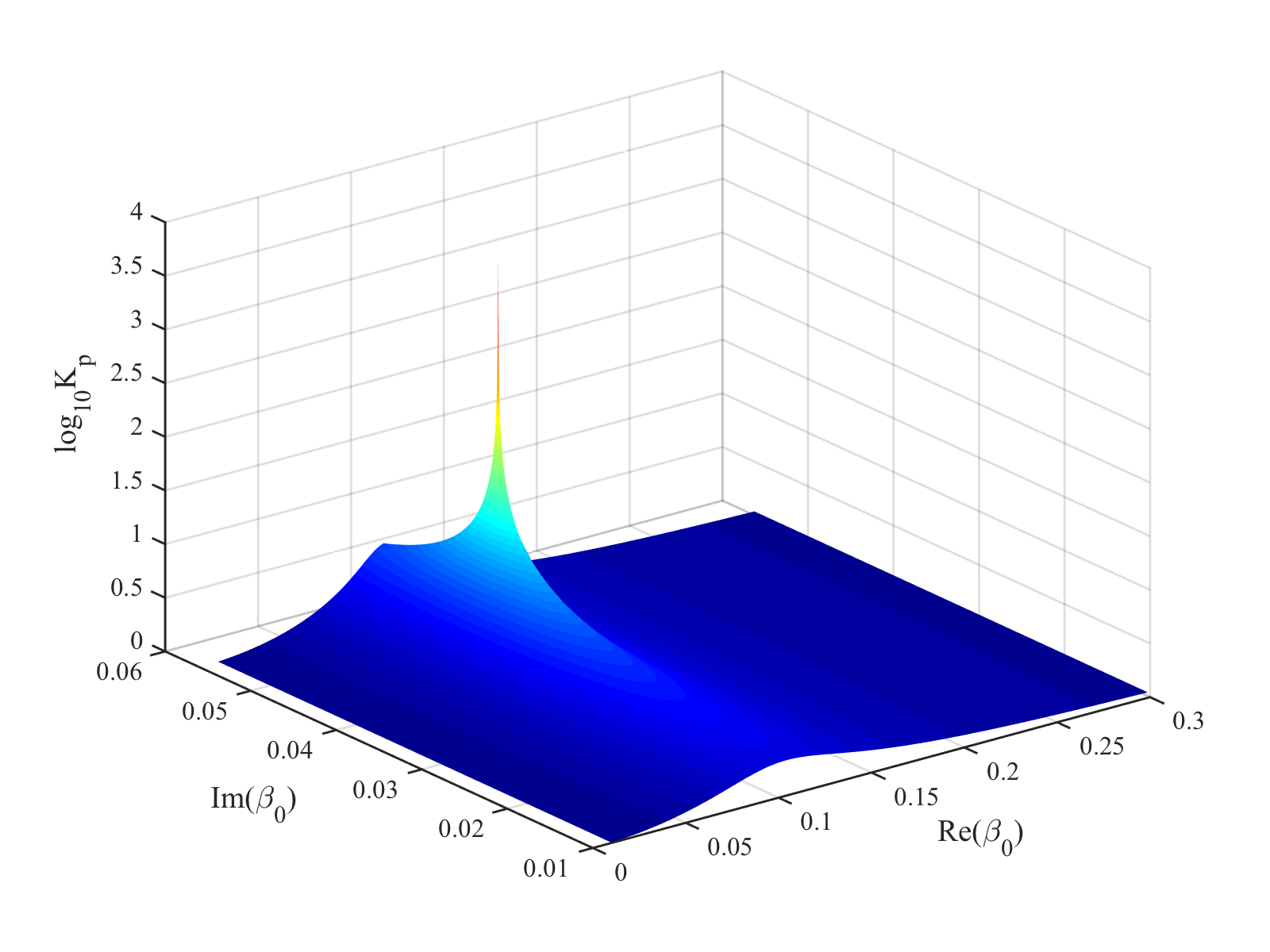}
\caption{\label{Kp} (Color online) Self-nonorthogonality log$_{10}(K_p)$ over the complex $\beta_0$ plane for the first mode $n=0$. $m=0$.} 
\end{figure}

 Case 2: At Cremer's optimum impedances or EPs. Eigenfunctions coalesce between a pair of neighbour modes; Left and right eigenfunctions are self-orthogonal $\int_s\tilde{\phi}_{n}\tilde{\varphi}^*_{n}ds=0$, therefore $K_p\to\infty$.  In Figs. \ref{Sij_fig} and \ref{Kp}, we plot $S_{ij}$ for $i=0$ and $j=1$, and $K_p$ for $n=0$ vary over the complex $\beta_0$ plane. It is shown clearly that at $\beta_{\text{Cremer}}$($\beta_{\text{EP}}$)$=0.099346+0.042653j$, $S_{01}=1$ and $K_{p,0}=6000$ and tend to be infinite. Varying along an arbitrary path in the complex $\beta_0$ plane, $K_p$ and $S_{ij}$ varies between the two opposite extremes.

\section{Effects of $K_p$ and $S_{ij}$ in sound power attenuation}\label{propagation}

\begin{figure}[ht]
\begin{center}
\includegraphics[scale=0.45]{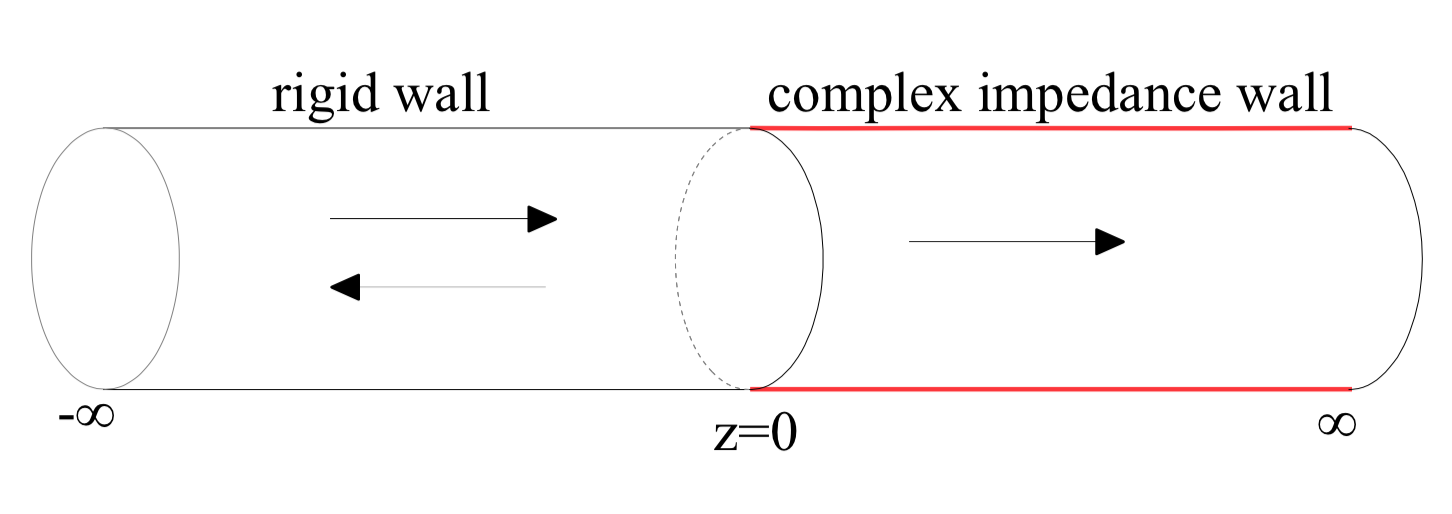}
\end{center}
\caption{\label{config} (Color online) Configuration of the example. The $z$-coordinate is aligned with the centre axis of the cylindrical waveguide. The left half semi-infinite wall ($z<0$) is rigid and the right half semi-infinite wall ($z>0$) is assumed as complex impedance boundary conditions.}
\end{figure}

In this section, we will give a simple example to illustrate the important roles of $K_p$ and $S_{ij}$ in sound power attenuation. As shown in Fig. \ref{config}, an infinite, cylindrical waveguide with circular cross-section is considered. The left half semi-infinite wall ($z<0$) is rigid and the right half semi-infinite wall ($z>0$) is assumed as complex uniform impedance boundary conditions.
 The sound pressure satisfies the Helmholtz equation 
\begin{equation}\label{helmholtz}
\nabla^2_\bot p+\frac{\partial^2p}{\partial z^2}+K^2p=0,
\end{equation}
where
\begin{equation}
\nabla^2_\bot =\frac{1}{r}\frac{\partial}{\partial
  r}(r\frac{\partial}{\partial
  r})+\frac{1}{r^2}\frac{\partial^2}{\partial \theta^2},
\end{equation}
and the boundary condition 
\begin{equation}\label{bound1}
\frac{\partial p}{\partial r}=Yp, \; \;  \mathrm{at} \; \; r=1,
\end{equation}
where $K=\omega R /c_0$ refers to the dimensionless wave number, $Y=-jK \beta_0$, and $\beta_0$ is the wall admittance.  Pressures and lengths are
respectively divided by $\rho_0 c_0^2$ and $R$ (the duct radius) and become dimensionless variables, where  $\rho_0$ and $c_0$ refer to ambient density and speed of sound in air, respectively. Rigid muilti-modes are incident from $-\infty$. Because the wall impedance is circumferentially uniform, eigenfunctions are decoupled in the circumferential direction. Without loss of generality, we consider only circumferential mode $m=0$.   

Sound pressure in the semi-infinite WIBC is expanded over right normalized eigenfunctions $\mathbf{\Phi}(r)=[\phi_0, \phi_1, \cdots, \phi_n, \cdots, \phi_N ]^T$
\begin{equation}\label{solution}
p(r, z)=\mathbf{\Phi}^T\mathsf{E}_l(z)\mathbf{C},
\end{equation}
where $\mathbf{C}$ is an amplitude vector of dimension $N$, $\mathsf{E}_l(z)$ is a diagonal matrix with $\exp(-jK_nz)$ on the main diagonal, with $K_n=\sqrt{K^2-\gamma_n^2}$. $\phi_n$ and $\gamma_n$ are defined in Eqs. (\ref{laplacian}, \ref{bound0}). $"^T"$ refers to transpose. $N$ refers to the truncation of the expansion. The eigenfunctions are normalized as defined in Eq. (\ref{normalize}). It is noted that although there is no mathematical theorem to guarantee the completeness of $\phi_n$, $n=0-\infty$, however, except at the infinite exceptional points, we found numerically that the expansion is convergent, in general, in the whole complex plane.  

The continuities of  pressure and axial particle velocity at $z=0$ lead to
\begin{align}
\mathbf{\Psi}^T(\mathbf{A} + \mathbf{B}) &= \mathbf{\Phi}^T \mathbf{C}, \label{eq_contin_p}\\
\mathbf{\Psi}^T\mathsf{K}_r( \mathbf{A} - \mathbf{B}) &= \mathbf{\Phi}^T\mathsf{K}_l \mathbf{C}, \label{eq_contin_v}
\end{align}
where $\mathbf{\Psi}=[\psi_0, \psi_1, \cdots, \psi_n, \cdots, \psi_N]^T$, $\mathsf{K}_r$ and $\mathsf{K}_l$ are diagonal matrices with the axial wavenumbers $\sqrt{K^2-\alpha_n^2}$  and $\sqrt{K^2-\gamma_n^2}$ on their main diagonal, respectively. $\psi_n$ and $\alpha_n$ are the normalised eigenfunctions and eigenvalues of modes in the semi-infinite waveguide with rigid boundary conditions. $\mathbf{A}$ and $\mathbf{B}$ are the amplitude vectors of incident and reflected modes.

Projecting Eq. (\ref{eq_contin_p}) over the left normalised eigenfunctions $\pmb\varphi^*=[\varphi_0, \varphi_1, \cdots, \varphi_n, \cdots, \varphi_N]^{*T}$, and Eq. (\ref{eq_contin_v}) over the normalised rigid eigenfunctions $\mathbf{\Psi}^*$, we obtain
\begin{align}
\mathsf{G} &=(\mathsf{K}_r+\mathsf{F}^T\mathsf{K}_l\mathsf{K}_p'\mathsf{F})^{-1}(\mathsf{K}_r-\mathsf{F}^T\mathsf{K}_l\mathsf{K}_p'\mathsf{F}),\nonumber\\
\mathbf{B} &=\mathsf{G}\mathbf{A},\\
\mathbf{C} &=\mathsf{K}_p'\mathsf{F}(\mathsf{I}+\mathsf{G})\mathbf{A},\nonumber
\end{align}
where 
\begin{equation}
\mathsf{F}=\int_0^1\pmb\varphi^*\mathbf{\Psi}^Trdr, 
\end{equation}
describes the couplings between the eigenfunctions of modes in the semi-infinite waveguides of rigid wall and complex impedance wall, respectively. $\mathsf{K}'_p$ is a diagonal matrix with $K'_{p,n}$ defined in Eq. (\ref{kp}) on the main diagonal.

Sound power in the semi-infinite WIBC is
\begin{align}\label{sp}
W &= \frac{1}{2}\Re e\{\int_0^1 p(r, \theta)v_z^*(r, \theta)rdr\}\\\nonumber
     &=\frac{1}{2}\Re e[\mathbf{A}^T(\mathsf{I}+\mathsf{G})^T\mathsf{F}\mathsf{E}_l(z)\mathsf{K}'_p\mathsf{S}_{ij}(\mathsf{K}_p')^*(\mathsf{K}_l/K)\mathsf{E}_l^*(z)\mathsf{F}^*(\mathsf{I}+\mathsf{G})^{*T}\mathbf{A}^*]\\\nonumber
     &=\underbrace{\frac{1}{2}\Re e[\sum_{i=j}\vert C_i'\vert^2K_{p,i}\frac{K_{l,i}^*}{K}e^{-\Im m(K_{l,i})z}]}_{\text{sum of sound power in individual mode}}+\underbrace{\frac{1}{2}\Re e[\sum_{i\ne j}C_i'C_j'^*K'_{p,i}S_{ij}(K_{p,j}')^*\frac{K_{l,j}^*}{K}e^{-j(K_{l,i}-K_{l,j}^*)z}]}_{\text{sum of cross power}},
\end{align}
where 
\begin{equation}
\mathsf{C}'=\mathsf{F}(\mathsf{I}+\mathsf{G})\mathbf{A},
\end{equation}
is a column vector, $C'_{i(j)}$ are the elements of $\mathsf{C}'$, $\mathsf{S}_{ij}$ is a matrix, its elements $S_{ij}$ are defined in Eq. (\ref{sij}), $K_{l,i(j)}$ are the elements $i(j)$ of the wavenumber matrix $\mathsf{K}_l$. 

Equation (\ref{sp}) clearly shows that the sound power are mainly decided not only by the individual mode attenuation factors $\mathsf{E}_l(z)$ (diagonal matrix with $e^{-jK_{l,i}z}$ on the main diagonal), but also by self-nonorthogonality $K_p$ and mutual-nonorthogonality $S_{ij}$, and $\mathsf{F}$ which describes the couplings between eigenfunctions of modes in the semi-infinite waveguides of rigid wall and complex impedance wall.  

\begin{figure}[ht]
\begin{center}
\includegraphics[scale=0.6]{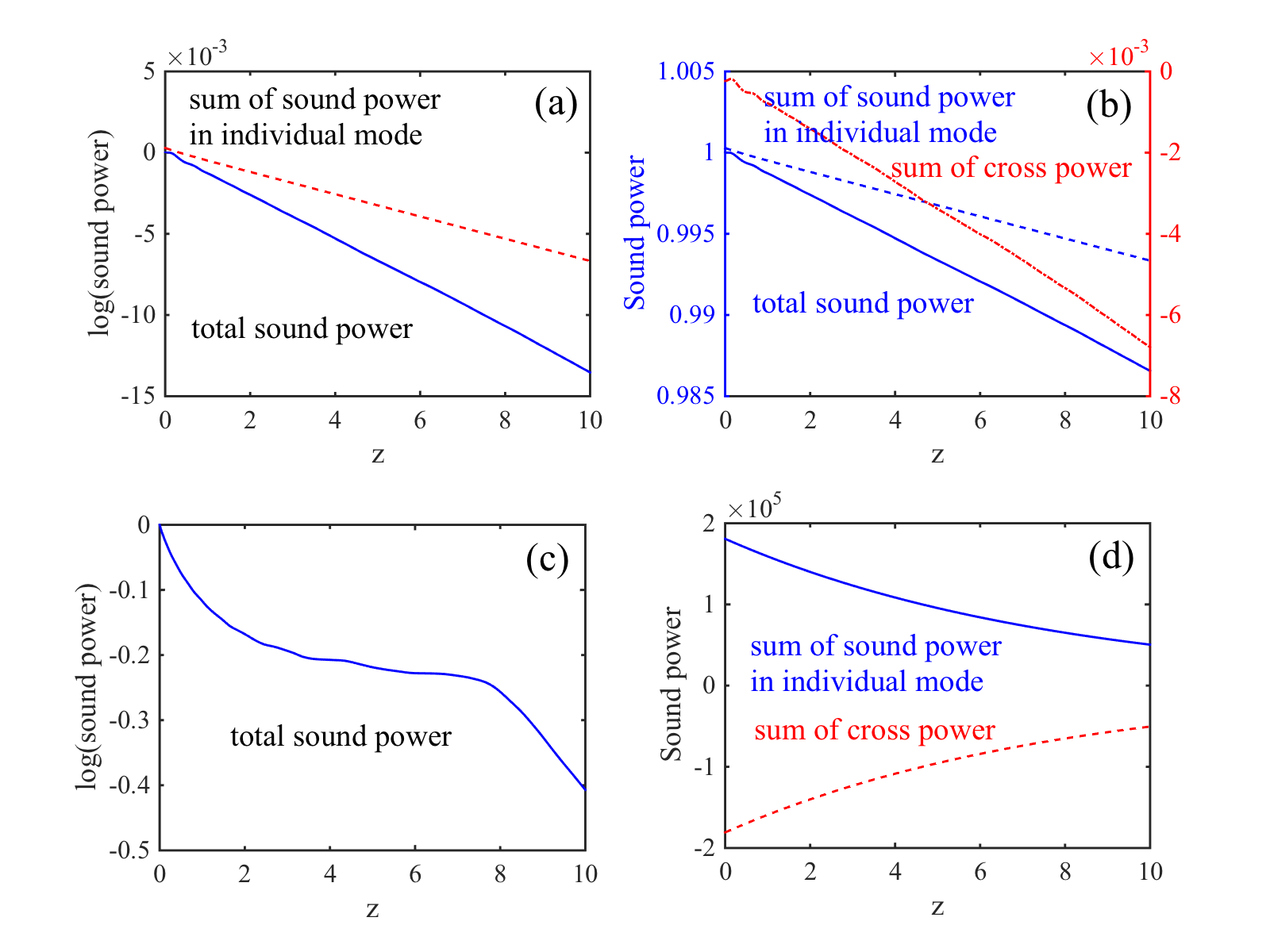}
\end{center}
\caption{\label{Wx_140815} (Color online) Sound power as a function of $z$.  (a) Total sound power (solid line) and sum of sound power in individual mode (dashed line) for case 1: $Z=0.1-j$. (Sound power is in logarithm scale.)  (b) Total sound power (solid line), sum of sound power in individual mode (dashed line) and sum of cross power (dash-dot line) for case 1: $Z=0.1-j$. (Sound power is in linear scale.) (c) Total sound power for case 2: $\beta_0=0.0993+0.0427j$. (Sound power is in logarithm scale.) (d) Sum of sound power in individual mode (solid line) and sum of cross power for case 2: $\beta_0=0.0993+0.0427j$. (Sound power is in linear scale.) $K=30$.}
\end{figure}

In Fig. \ref{Wx_140815}, we show sound power as a function of $z$ for two two cases: Case 1, $Z=0.1-j$ in which the dissipation in the boundary wall is less important; Case 2, $\beta_0=0.0993+0.0427$ which is close to the first EP ($m=0$), the boundary wall is very dissipative. Rigid multi-modes with coefficient
\begin{figure}[ht]
\begin{center}
\includegraphics[scale=0.6]{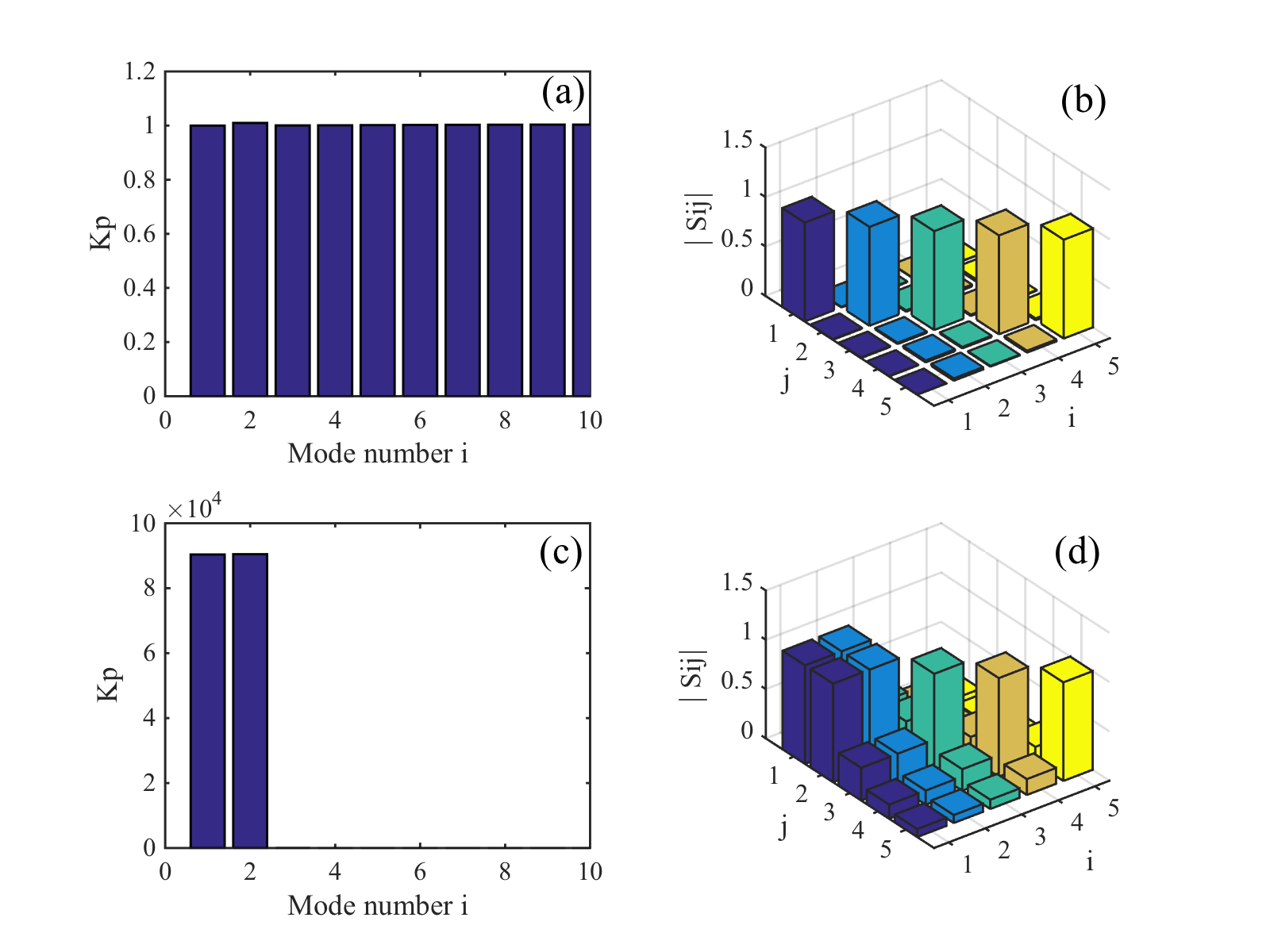}
\end{center}
\caption{\label{Kp_Sij_140815} (Color online) Self-nonorthogonality $K_p$ and mutual-nonorthogonality $S_{ij}$. (a) $K_p$ for case 1, $Z=0.1-j$. (b) $S_{ij}$ for case 1, $Z=0.1-j$. (c) $K_p$ for case 2, $\beta_0=0.0993+0.0427j$. $S_{ij}$ for case 2, $\beta_0=0.0993+0.0427j$. $K=30$.}
\end{figure}
\begin{equation}
A_i=\int_0^1\psi_i\varphi_0rdr=\int_0^1\psi_i\frac{1}{\sqrt{\Lambda_0}}(\frac{J_0(\gamma_0r)}{J_0(\gamma_0)})^*rdr,
\end{equation}
are incident, where ``${}^*$" refers to complex conjugate, $i=0-50$ are incident rigid mode index. $J_0$ is zero order Bessel function. Without loss of generality, we use $K=30$, which is a typically industrial value in the lined intakes of an aeroengine. 

For case 1, small dissipation is included, $K_p$ are approximately equal to $1$ for all modes as shown in Fig. \ref{Kp_Sij_140815} (a). Modes are approximately mutual-orthogonal. $S_{ij}$ ($i\ne j$) are approximately equal to zero as shown in Fig. \ref{Kp_Sij_140815} (b). The total sound power and the sum of sound power in individual mode decrease exponentially, as we expect intuitively, as shown in Fig. \ref{Wx_140815} (a). The sum of cross-power is not important as shown in Fig. \ref{Wx_140815} (b). (For comparison, we plot also in Fig. \ref{Wx_140815} (b) the total sound power and the sum of sound power in individual mode. Note that they are shown in linear scale.) This conclusion can be also obtained directly from Eq. (\ref{sp}). 

However, for case 2 in which the impedance is close to the first EP, the acoustically absorbent material is very dissipative, it is very surprise that the total sound power attenuation curve has a plateau between about $z=2$ and $z=8$ where total sound power almost does not attenuate as shown in Fig. \ref{Wx_140815} (c). The sum of sound power in individual mode decreases still exponentially (not shown), but with very large amplitude (about $10^5$) as shown in Fig. \ref{Wx_140815} (d). The sum of cross-power are negative and increases with $z$ with almost the same order of amplitude. By these results, we can conclude safely that although Cremer's optimum impedances give the maximum attenuation for individual mode, however, they stimulate simultaneously very high amplitudes for the corresponding mode, the total sound power attenuation is mainly decided by the sum of sound power in individual mode and the sum of cross power cancel each other out.   

This can be explained also by Eq. (\ref{sp}) in which the amplitudes of modes $i=0$ and $j=1$ are extremely larger than these of other modes. The two modes dominate the total sound power, the sum of sound power in individual mode, and the sum of cross power over other modes. The extremely large amplitudes are due to the nearly self-orthogonalities between left and right eigenfunctions of the almost coalescent modes $i=0$ and $j=1$ near the first EP. This can be seen by the $K_p$ shown in Fig. \ref{Kp_Sij_140815} (c) whose values are approximately equal to $9\times 10^4$ for modes $i=0$ and $j=1$ and are extremely larger than these of other modes. The important effects of cross-power are produced by the mutual-nonorthogonality $S_{ij}$. In this case, $S_{01}$ and $S_{10}$ are approximately equal to $1$ as shown in Fig. \ref{Kp_Sij_140815} (d). Because the two modes $i=0$ and $i=1$ are almost coalescent, $K'_{p,0}$ and $K'_{p,1}$ are approximately equal. Similary, $C_0'$, $K_0$ are are approximately equal to $C'_1$ and $K_1$, therefore the sum of sound power in individual mode and the sum of cross power cancel each other out.

\section{Conclusions}

In this paper, we have given new insights into the nonorthogonalities of eigenfunctions in a waveguide with impedance boundary conditions. We have defined two quantities: self-nonorthogonality $Kp$, which measures the nonorthogonality between left and right eigenfunctions of a mode, and mutual-nonorthogonality $S_{ij}$, which measures the nonorthogonality between modes $i$ and $j$. Two opposite limiting cases are clearly identified in the complex boundary admittance $\beta_0$ plane. One is non-dissipation, i.e., acoustic rigid, pressure-release, and purely reactive impedance. Modes are mutual-orthogonal. Left eigenfunctions are equal to right eigenfunctions. $S_{ij}=0$ and $K_p=1$. The other is Cremer's optimum impedances which are exceptional points. Both eigenvalues and eigenfunctions coalesce between a pair of neighbour modes $i$ and $j$. Left and right eigenfunctions of the coalescent modes are self-orthogonal. In this case, $S_{ij}=1$ and $K_p=\infty$. 

The total sound power in the waveguide is mainly decided by $K_p$, $S_{ij}$, besides the exponential attenuation factors $e^{-jK_lz}$. We have shown that although Cremer's optimum impedances give the maximum attenuation for individual mode, however, they simultaneously stimulate very high amplitudes for the corresponding mode. When the acoustically absorbent materials are very dissipative, the total sound power attenuation is mainly decided by the cancel each other out between the sum of sound power in individual mode and the sum of cross power.

\appendix

\section{\label{appendix-I} Adjoint eigenvalue problem and bi-orthogonal relation}

We consider an operator $\mathscr{L}=\nabla^2_\bot+\gamma_{mn}^2$, the eigenvalue problem defined by Eqs. (\ref{laplacian}) and (\ref{bound0}) can be rewritten as
\begin{equation}\label{app1}
\mathscr{L}\tilde{\phi}_{mn}=0,
\end{equation}
with the boundary condition
\begin{equation}\label{app3}
\mathscr{G}\tilde{\phi}_{mn}=0,\; \text{at}\; r=1,
\end{equation}
where $\mathscr{G}=\partial/\partial r-Y$. By introducing a function $\tilde{\varphi}$ which satisfies 
\begin{align}
\int_s\tilde{\varphi}^*_{m'n'}\mathscr{L}\tilde{\phi}_{mn}ds & =\int_s\tilde{\varphi}^*_{m'n'}(\nabla^2_\bot+\gamma_{mn}^2)\tilde{\phi}_{mn}ds\\
& =\int_s\tilde{\phi}_{mn}[\nabla^2_\bot+\gamma_{mn}^2]\tilde{\varphi}^*_{m'n'}ds+\oint\left (\tilde{\varphi}^*_{m'n'}\frac{\partial\tilde{\phi}_{mn}}{\partial r}-\tilde{\phi}_{mn}\frac{\partial\tilde{\varphi}^*_{m'n'}}{\partial r}\right)dC\\\nonumber
&=\int_s(\mathscr{L}^+\tilde{\varphi}_{m'n'})^*\tilde{\phi}_{mn}ds-\int_0^{2\pi}\tilde{\phi}_{mn}\left (\frac{\partial\tilde{\varphi}_{m'n'}^*}{\partial r}-Y\tilde{\varphi}^*_{m'n'}\right)d\theta\\\nonumber
& =\int_s(\mathscr{L}^+\tilde{\varphi}_{m'n'})^*\tilde{\phi}_{mn}ds-\int_0^{2\pi}\tilde{\phi}_{mn}(\mathscr{G}^+\tilde{\varphi}_{m'n'})^*,\nonumber
\end{align}
where
\begin{align}
\mathscr{L}^+=\nabla^2_\bot+(\gamma_{mn}^2)^*=\nabla^2_\bot+\epsilon_{m'n'}^2, \;\;\;\;  \mathscr{G}^+=\partial/\partial r-Y^*,
\end{align}
we can define the adjoint eigenvalue problem as
\begin{equation}\label{app2}
\mathscr{L}^+\tilde{\varphi}_{m'n'}=0,
\end{equation}
with the boundary condition
\begin{equation}\label{app4}
 \mathscr{G}^+\tilde{\varphi}_{m'n'}=0,\; \text{at}\; r=1.
\end{equation}

Multiplying Eq. (\ref{app1}) by $\tilde{\varphi}^*_{m'n'}$ and the conjugate of Eq. (\ref{app2}) by $\tilde{\phi}_{mn}$ for different mode $\gamma^2_{mn}\ne (\epsilon^2_{m'n'})^*$, i.e., $m\ne m'$ and $n\ne n'$, and subtracting the results, we obtain
\begin{align}
[\gamma_{mn}^2-(\epsilon_{m'n'}^2)^*]\int_s\tilde{\phi}_{mn}\tilde{\varphi}_{m'n'}^*ds &=\int_s(\tilde{\phi}_{mn}\nabla^2_\bot\tilde{\varphi}^*_{m'n'}-\tilde{\varphi}_{m'n'}^*\nabla^2_\bot\tilde{\phi}_{mn})ds\\
& =\oint\left (\tilde{\phi}_{mn}\frac{\partial\tilde{\varphi}^*_{m'n'}}{\partial r}-\tilde{\varphi}^*_{m'n'}\frac{\partial\tilde{\phi}_{mn}}{\partial r}\right)dC\\\nonumber
& = \int_0^{2\pi}\tilde{\phi}_{mn}\left (\frac{\partial\tilde{\varphi}^*_{m'n'}}{\partial r}-Y\tilde{\varphi}^*_{m'n'}\right)d\theta\\\nonumber
&=0,
\end{align}
where we have used Eq. (\ref{app4}), $\mathscr{G}^+\tilde{\varphi}_{mn}=0$. Therefore,
\begin{equation}\label{ortho_app}
\int_s\tilde{\phi}_{mn}\tilde{\varphi}^*_{m'n'}ds=0,\;\; \text{when}\; m\ne m',\; n\ne n'.
\end{equation}

When the boundary is acoustically rigid ($Y=0$), pressure release ($Z=0$), or purely reactive without dissipation ($\beta_0=jc$, $Y=-jK\beta_0=Kc$, $c$ is real),  
\begin{equation}
\mathscr{L}^+=\mathscr{L}, \;\;   \mathscr{G}^+= \mathscr{G},
\end{equation}
the eigenvalue problem defined by Eqs. (\ref{laplacian}) and (\ref{bound0}) or (\ref{app1}) and (\ref{app3}) is self-adjoint. It is easy to show that $\gamma_{mn}^2$ are real and $\tilde{\varphi}_{mn}=\tilde{\phi}_{mn}$. 

On the other hand, for all practical problem, $\beta_0$ and therefore $Y$ are complex. Taking the complex conjugate of Eqs. (\ref{app2}) and (\ref{app4}), we obtain
\begin{equation}
\tilde{\varphi}_{mn}=\tilde{\phi}_{mn}^*.
\end{equation}  
The orthogonal relation (\ref{ortho_app}) is then rewritten as the bi-orthogonality (\ref{bi-orthogonal})
\begin{equation}\label{bi_ortho_app}
\int_s\tilde{\phi}_{mn}\tilde{\varphi}_{m'n'}^*ds=\int_s\tilde{\phi}_{mn}\tilde{\phi}_{m'n'}ds=0,\;\; \text{when}\; m\ne m',\; n\ne n'.
\end{equation}

\section{Branch point on the complex admittance plane}

We consider the variation of eigenvalues as a function of complex admittance. At some admittances, the eigenvalues of a pair of neighbour modes coalesce to form Branch Points (BP) on the complex admittance plane. At the BP, the dispersion equation (\ref{dispersion1}) has double eigenvalues $\gamma_{\text{BP}}$, i.e.
\begin{align}\label{double_root}
\gamma_{m_0n}\frac{J'_{m_0}(\gamma_{m_0n})}{J_{m_0}(\gamma_{m_0n})}\Bigm\vert_{\gamma_{m_0n}=\gamma_{\text{BP}}} &=-jK\beta_{\text{BP}},\\
\frac{\partial}{\partial\gamma_{m_0n}}\left(\gamma_{m_0n}\frac{J'_{m_0}(\gamma_{m_0n})} {J_{m_0}(\gamma_{m_0n})}\right)\Bigm|_{\gamma_{m_0n}=\gamma_{\text{BP}}} &=0,
\end{align} 
where we have fixed $m=m_0$, $\gamma_{m_0n}$ are eigenvalues, $J_{m_0}$ are the $m_0$ order Bessel function, the prime refers to derivative with respect to $\gamma_{m_0n}$. $K$ is dimensionless frequency. If we define a function 
\begin{equation}
f(\gamma_{m_0n}, \beta_0)=\gamma_{m_0n}\frac{J'_{m_0}(\gamma_{m_0n})}{J_{m_0}(\gamma_{m_0n})} + jK\beta_0,
\end{equation}
we expand the function $f(\gamma_{m_0n}, \beta_0)$ in the vicinity of $(\gamma_{\text{BP}}, \beta_{\text{BP}})$ to the lowest order as
\begin{equation}
f(\gamma_{m_0n}, \beta_0)\approx\frac{1}{2}\frac{\partial^2 f}{\partial\gamma^2_{m_0n}}\Bigm\vert_{\gamma_{m_0n}=\gamma_{\text{BP}}}(\gamma_{m_0n}-\gamma_{\text{BP}})^2+\frac{\partial f}{\partial\beta_0}\Bigm\vert_{\beta_0=\beta_{\text{BP}}}(\beta_0-\beta_{\text{BP}}).
\end{equation}
Suppose that in the vicinity of $(\gamma_{\text{BP}}, \beta_{\text{BP}})$, $\partial^2 f/\partial\gamma^2_{m_0n}\ne 0$, i.e. there has no triple or higher order eigenvalues of dispersion equation (\ref{dispersion1}), we obtain,
\begin{equation}
\gamma_{m_0n}-\gamma_{\text{BP}}\approx-\sqrt{\frac{2\partial f/\partial\beta_0}{\partial^2 f/\partial\gamma^2_{m_0n}}}\sqrt{\beta_0-\beta_{\text{BP}}}.
\end{equation}
$\beta_{\text{BP}}$ is a square root branch point in the complex admittance plane.

%\section*{Appendix II:  $(\phi^l)^*:\phi^r$ (P39)}

%Appendix III: The completeness of liner modes

%Appendix : the independent of modes


\begin{thebibliography}{99}
%{References and links}
%\bibliography
%\textbf{References and links}\\

\bibitem{pierce} A. D. Pierce, \textit{Acoustics, An introduction to its physical principles and applications} (McGraw-Hill Book Company, New York, 1981), Chap. VII.

\bibitem{morse} P. M. Morse and K. U. Ingard, \textit{Theoretical acoustics} (McGraw-Hill Book Company, New York, 1968), Chap. IX.

\bibitem{nayfeh}  A.\ H.\ Nayfeh, J.\ E.\ Kaiser and D.\ P.\ Telionis, ``Acoustics of aircraft engine-duct systems," AIAA J. \textbf{13}, 130-153 (1975).

%\bibitem{kraft} R. E. Kraft, W. R. Wells, Adjointness properties for differential systems with eigenvalue-dependent boundary conditions, with application to flow-duct acoustics, \textit{J. Acoust. Soc. Am. }61, 913-922, 1977.

\bibitem{cremer} L. Cremer, ``Theory of sound attenuation in a rectangular duct with an absorbing wall and the resultant maximum attenuation coefficient," (in german) Acustica \textbf{2}, 249-263 (1953). 

\bibitem{tester1}  B.\ J. Tester, ``The optimization of modal sound attenuation in duct, in the absence of mean flow,"  J. Sound Vib. \textbf{27}, 477-513 (1973).

\bibitem{zorumski} W. E. Zorumski and J. P. Mason, ``Multiple eigenvalues of sound-absorbing circular and annular ducts," J. Acoust. Soc. Am. \textbf{55}, 1158-1165 (1974).

\bibitem{eversman} W. Eversman, \textit{Theoretical models for duct acoustic propagation and radiation}, in \textit{Aeroacoustics of flight vehicles: theory and practice. Volume 2: noise control}, AD-A241, 142 (1991), Chap. XIII.

\bibitem{mechel1}  F.\ P. Mechel, ``Modal solutions in rectangular ducts lined with locally reacting absorbers," Acustica \textbf{73}, 223-239 (1991).

\bibitem{mechel2}  F.\ P. Mechel, ``Modal solutions in circular and annular ducts with locally or bulk reacting lining," Acustica \textbf{84}, 201-222 (1998).

\bibitem{koch} W. Koch, ``Attenuation of sound in multi-element acoustically lined rectangular ducts in the absence of mean flow," J. Sound Vib. \textbf{52}, 459-496, (1977).
 
\bibitem{rice0} E. J. Rice, ``Multimodal far-field acoustic radiation pattern using mode cutoff ratio," AIAA J. \textbf{16}, 906-911 (1978).
 
\bibitem{rice1} E. J. Rice, ``Optimum wall impedance for spinning modes - a correlation with mode cut-off ratio," J. Aircraft \textbf{16}, 336-343 (1979).

\bibitem{rice2} L. J. Heidelberg and E. J. Rice, ``Experimental evaluation of a spinning-mode acoustic treatment," NASA-1613 (1980). 

\bibitem{eversman1} W. Eversman, ``Effect of lining non-linearity on realized attenuation of tonal noise," Procedia Engineering \textbf{6}, 114-123 (2010). 

\bibitem{watson} W. R. Watson, ``Circumferentially segmnted duct liners optimized for axisymmetric and standind-wave sources,"  NASA-2075 (1982).

\bibitem{watson1} W. R. Watson, M. G. Jones, T. L. Parrott, and J. Sobieski, ``Assessment of equation solvers and optimization techniques for nonaxisymmetric liners," AIAA J. \textbf{42}, 2010-2018 (2004).
 
\bibitem{campos} L. M. B. C. Campos and J. M. G. S. Oliveira, ``On the acoustic modes in a cylindrical duct with an arbitrary wall impedance distribution," J. Acoust. Soc. Am. \textbf{116}, 3336-3347 (2004).
 
\bibitem{bielak} G. W. Bielak, J. W. Premo and A. S. Hersh, ``Advanced turbofan duct liner concepts," NASA/CR-1999-209002 (1999).

\bibitem{morse1} P. M. Morse, ``The transmission of sond pipes," J. Acoust. Soc. Am. \textbf{11}, 205-210 (1939).

\bibitem{shenderov} E. L. Shenderov, ``Helmholtz equation solutions corresponding to multiple roots of the dispersion equation for a waveguide with impedance walls," Acoustical Physics \textbf{46}, 357-363 (2000).

\bibitem{rienstra} S. W. Rienstra, ``A classification of duct modes based on surface waves," Wave Motion \textbf{37}, 119-135 (2003).   

\bibitem{dembowski1} C. Dembowski, H. D. Graf, H. L. Harney, A. Heine, W. D. Heiss, H. Rehfeld, and A. Richter, ``Experimental observation of the topological structure of exceptional points," Phys. Rev. Lett. \textbf{86}, 787 (2001).

\bibitem{heiss1} W. D. Heiss and A. L. Sannino, ``Avoided level crossing and exceptional points," J. Phys. A: Math. Gen. \textbf{23}, 1167-1178 (1990).

\bibitem{heiss2} W. D. Heiss and A. L. Sannino, ``Transitional regions of finite Fermi systems and quantum chaos," Phys. Rev. A \textbf{43}, 4159-4166 (1991).

\bibitem{heiss4} W. D. Heiss, ``Exceptional points of non-Hermitian operators," J. Phys. A: Math. Gen. \textbf{37}, 2455-2464 (2004).  
  
\bibitem{heiss3} W. D. Heiss, ``The physics of exceptional points," J. Phys. A: Math. Theor. \textbf{45}, 444016 (2012).

\bibitem{rotter} I. Rotter, ``A non-Hermitian Hamilton operator and the physics of open quantum systems," J. Phys. A: Math. Theor. \textbf{42}, 153001 (2009).

\bibitem{berry} M. V. Berry, ``Physics of nonhermitian degeneracies," Czechoslovak J. of Phys. \textbf{54}, 1039-1047 (2004).

%\bibitem{kato} T. Kato, Perturbation Theory of Linear Operators, Berlin: Springer, 1966.

\bibitem{latinne} O. Latinne, N. J. Kylstra, M. Dšrr, J. Purvis, M. Terao-Dunseath, C. J. Joachain, P. G. Burke, and C. J. Noble, ``Laser-induced degeneracies involving autoionizing states in complex atoms," Phys. Rev. Lett. \textbf{74}, 46 (1995).

%\bibitem{Oberthaler} M. K. Oberthaler, R. Abfalterer, S. Bernet, J. Schmiedmayer, and A. Zeilinger, Atom Waves in Crystals of Light, Phys. Rev. Lett. \textbf{77}, 4980 (1996).

\bibitem{Stehmann} T. Stehmann, W. D. Heiss and F. G. Scholtz, ``Observation of exceptional points in electronic circuits," J. Phys. A \textbf{37}, 7813 (2004).

\bibitem{cartarius} H. Cartarius, J. Main, and G. Wunner, ``Exceptional points in atomic spectra," Phys. Rev. Lett. \textbf{99}, 173003 (2007).

%\bibitem{Dembowski2} C. Dembowski, B. Dietz, H.-D. GrŠf, H. L. Harney, A. Heine, W. D. Heiss, and A. Richter, Observation of a Chiral State in a Microwave Cavity, Phys. Rev. Lett. \textbf{90}, 034101 (2003).  

%\bibitem{Dembowski3} C. Dembowski, B. Dietz, T. Friedrich, H.-D. GrŠf, A. Heine, C. Mej'a-Monasterio, M. Miski-Oglu, A. Richter, and T. H. Seligman, First Experimental Evidence for Quantum Echoes in Scattering Systems, Phys. Rev. Lett. \textbf{93}, 134102 (2004).

\bibitem{lee} S. B. Lee, J. Yang, S. Moon, S-Y Lee, J-B Shim, S. W. Kim, J-H Lee, and K. An, ``Observation of an exceptional point in a chaotic optical microcavity," Phys. Rev. Lett. \textbf{103}, 134101 (2009).

%\bibitem{seyrain} A. P. Seyranian \textit{et al.}, J. Phys. A: Math. Gen. \textbf{38} 1723 (2005).

\bibitem{moiseyev} S. Klaiman, U. G\"{u}nther, and N. Moiseyev, ``Visualization of branch points in PT-symmetric waveguides," Phys. Rev. Lett. \textbf{101}, 080402 (2008).

\bibitem{bi_resonance_trap} W. P. Bi and V. Pagneux, ``Resonance trapping in waveguides with impedance boundary conditions," to be submitted.

\bibitem{leissa} A. W. Leissa, ``On a curve veering aberration," J. Applied Math. and Phys. (ZAMP) \textbf{25}, 99-111 (1974).

\bibitem{kuttler} J. R. Kuttler and V. G. Sigillito, ``On curve veering," J. Sound and Vib. \textbf{75}, 585-588 (1981).


%*******************************************************

%\bibitem{lansing}  D.\ L.\ Lansing \& W.\ E. Zorumski, Effects of wall admittance changes on duct transmission and radiation of sound. \textit{J. Sound Vib.} 27, 85-100, 1973.

%\bibitem{unruh}  J.\ F. Unruh, Finite length tuning for low frequency lining design. \textit{J. Sound Vib.} 45, 5-14, 1976.

%\bibitem{beckemeyer1}  R.\ J.\ Beckemeyer \& D.\ Sawdy, Boundary conditions for mode-matching analyses of coupled acoustic fields in ducts. \textit{AIAA Journal, }16, 912-918, 1978.

%\bibitem{joshi} M. C. Joshi, R. E. Kraft, G. fiske, A. A. syed and R. E. Motsinger, Sound propagation in segmented exaust ducts - theoretical predictions and comparison with measurements, \textit{AIAA-83-0734}, 1983.

%\bibitem{eversman10}  W. Eversman \& E.\ L. Cook, A method of weighted residuals for the investigation of sound transmission in non-uniform ducts without flow. \textit{J. Sound Vib.} 38,105-123, 1975.

%\bibitem{eversman11}  W. Eversman \& R.\ J. Astley, Acoustic transmission in non-uniform ducts with mean flow, part I: the method of weighted residuals. \textit{J. Sound Vib.} 74, 89-101, 1981.

%\bibitem{dougherty1} R. P. Dougherty, A wave-splitting technique for nacelle acoustic propagation, \textit{AIAA-97-1652}, 1997.

%\bibitem{dougherty2} R. P. Dougherty, A parabolic approximation for flow effects on sound propagation in nonuniform, softwall, ducts, \textit{AIAA-99-1822}, 1999.

%\bibitem{lan} J. H. Lan, Turbofan duct propagation model, \textit{NASA/CR-2001-211245}, 2001.

%\bibitem{rienstra1}  S.\ W. Rienstra 1999 Sound transmission in slowly varying circular and annular lined ducts with flow. \textit{J. Fluid Mech. } 380, 279-296, 1999.

%\bibitem{rienstra2} S. W. Rienstra, W. Eversman, A numerical comparison between the multi-scales and finite-element solution for sound propagation in lined flow ducts, \textit{J. Fluid Mech. } 437, 367-384, 2001.

%\bibitem{dougherty} R. P. Dougherty, Nacelle acoustic design by ray tracing in three dimensions, \textit{AIAA-96-1773}, 1996.

%\bibitem{cummings} A. Cummings, High frequency ray acoustics models for duct silencers, \textit{J. Sound Vib.} 221, 681-708, 1999.

%\bibitem{dunn1} M. H. Dunn, TBIEM3D - A computer program for predicting ducted fan engine noise, version 1.1, \textit{NASA/CR-97-206232}, 1997.

%\bibitem{dunn2} M. H. Dunn, J. Tweed, F. Farassat, The application of a boundary integral equation method  to the prediction of ducted fan engine noise, \textit{J. Sound Vib.} 227, 1019-1048, 1999.

%\bibitem{lidoine} S. Lidoine, H. Batard, S. Troyers, A. Delnevo, M. Roger, Acoustic radiation modelling of aero-engine intake comparison between analytical and numerical methods, \textit{AIAA 2001-2140}, 2001.

%\bibitem{eversman14} W. Eversman, The boundary condition at an impedance wall in a non-uniform duct with potential mean flow, \textit{J.Sound Vib.} 246, 63-69, 2001.

%\bibitem{mcalpine1} A. McAlpine, M. C. M. Wright, H. Batard, S. Thezelais, Finite/voundary element assessment of a turbofan splices intake liner at supersonic fan operating conditions. \textit{AIAA 2003-3305}, 2003.

%\bibitem{long} Y. Ozyoruk, V. Ahuja, L. N. Long, Time domain simulation of radiation from ducted fans with liners, \textit{AIAA 2001-2171}, 2001.

%\bibitem{parrott} T. L. Parrott, M. G. Jones, Parallel-element liner impedances for improved absorption of broadband sound in ducts, \textit{Noise Control Eng. J.} 43, 183-195, 1995.

%\bibitem{sobolev} A. F. Sobolev, N. M. Solov'eva, \& R. D. Filippova, A way to expand the frequency range of sound absorbent liners for aircraft power plants, \textit{Acoustical physics}, 41, 124-129, 1995.

%\bibitem{baumeister} K. J. Baumeister, Evaluation of optimized multisectioned acoustic liners. \textit{AIAA J.} 17, 1185-1192, 1979.

%\bibitem{koch1}  W. Koch, Radiation of sound from a two-dimensional acoustically lined duct. \textit{J. Sound Vib. }55, 255-274, 1977.

%\bibitem{fuller} C. R. Fuller, Propagation and radiation of sound from flanged circular ducts with circumferentially varying wall admittances, I: semi-infinite ducts, \textit{J. Sound Vib.} 93, 321-340, 1984.

%\bibitem{fuller2} C. R. Fuller, Propagation and radiation of sound from flanged circular ducts with circumferentially varying wall admittances, II: finite ducts with sources, \textit{J. Sound Vib.} 93, 341-351, 1984.

%\bibitem{mani}  R. Mani, Acoustic duct with peripherally segmented acoustic treatment, \textit{United states patent No. 3,937,590,} 1976.

%\bibitem{howe}  M.\ S. Howe, The attenuation of sound in a randomly lined duct. \textit{J. Sound Vib.} 87, 83-103, 1983.

%\bibitem{sarin} S. L. Sarin, E. R. Rademaker, In flight acoustic mode measurements in the turbofan engine inlet of Fokker 100 aircraft, \textit{AIAA-paper 93-4414}.

%\bibitem{rademaker} E. R. Rademaker, S. L. Sarin, and C. A. Parente, Experimental investigation on the influence of liner non-uniformities on prevailing modes, \textit{AIAA 96-1682}.

%\bibitem{regan}  B. Regan \& J. Eaton, Modeling the influence of acoustic liner non-uniformities on duct modes, \textit{J. Sound Vib.} 219, 859-879, 1999.

%\bibitem{mcalpine3} A. McAlpine, M. C. M. Wright, H. Batard, S. Thezelais, Finite/voundary element assessment of a turbofan splices intake liner at supersonic fan operating conditions. \textit{AIAA 2003-3305}.

%\bibitem{groeneweg} H. F. Groeneweg, T. G. Sofrin, E. J. Rice \& Phillip R. Gliebe, Turbomachinery noise, in Aeroacoustics of flight vehicles: theory and practice - volume 1: noise sources, \textit{ADA 241 141} 151-209.



%\bibitem{wiersig} J. Wiersig, Formation of Long-Lived, Scarlike Modes near Avoided Resonance Crossings in Optical Microcavities, Phys. Rev. Lett. \textbf{97}, 253901(2006).

\bibitem{hodges} C. H. Hodges, ``Confinement of vibration by structural irregularity," J. Sound Vib. \textbf{82}, 441-424 (1982).

\bibitem{hodges1} C. H. Hodges and J. Woodhouse, ``Vibration isolation from irregularity in a nearly periodic structure: theory and measurements,"  J. Acoust. Soc. Am. \textbf{74}, 894-905 (1983).

\bibitem{pierre} C. Pierre and E. H. Dowell, ``Localization of vibrations by structural irregularity," J. Sound Vib. \textbf{114}, 549-564 (1987).

\bibitem{pierre1} C. Pierre, ``Mode localization and eigenvalue loci veering phenomena in disordered structures," J. Sound Vib. \textbf{126}, 485-502 (1988).

\bibitem{triantafyllou} M. S. Triantafyllou and G. S. Triantafyllou, ``Frequency coalescence and mode localization phenomena: a geometric theory," J. Sound Vib. \textbf{150}, 485-500 (1991).

\bibitem{petermann} K. Petermann, ``Calculated spontaneous emission factor for double-heterostructure injection lasers with gain-induced waveguiding," IEEE J. Quantum Electron. \textbf{QE-15} 566 (1979).


%\bibitem{persson} E. Persson, I. Rotter, H.-J. Stšckmann, and M. Barth, Observation of Resonance Trapping in an Open Microwave Cavity, Phys. Rev. Lett. 85, 2478 (2000).

%\bibitem{dembowski3} C. Dembowski, B. Dietz, H. D. Graf, H. L. Harney, A. Heine, W. D. Heiss, and A. Richter, Encircling an exceptional point,  Phys. Rev. E \textbf{69}, 056216 (2004).

%\bibitem{rice} E. J. Rice, Inlet noise suppressor design method based on the distribution of acoustic power with mode cut-off ratio. 1976 NASA CP-2001.

%\bibitem{berry1} M. V. Berry, Mode degeneracies and the Petermann excess-noise factor for unstable lasers, J. Mordern optics, \textbf{50}, 63-81, 2003.




%\bibitem{brambley}


\end{thebibliography}
\end{document}